# A Continuum Theory of Dynamically Loaded Polymers

B. E. Clements

Theoretical Division, Los Alamos National Laboratory, Los Alamos, NM 87545

bclements@lanl.gov

## Abstract

A thermo-mechanical continuum theory is proposed for dynamically loaded glassy polymers. The theory is based on an *ansatz* for the Helmholtz free energy where both the deviatoric and the volumetric contributions to the free energy are rate-dependent. The requirement that the free energy is fully rate dependent arises from the need to model the full range of conditions between those found in quasi-static applications to those common in high-rate shock loading scenarios. Using a purely equilibrium equation of state is found to be insufficient. The resulting model, called the Glassy Amorphous Polymer (GAP) model, suitably captures the thermo-mechanical behavior of both equilibrium properties, but also high-rate phenomena like the shock Hugoniot. Polymethylmethacrylate (PMMA) is used as a representative polymer because it is one of the few polymers where sufficient experiments have been done to determine many of the parameters required by the GAP model. An important part of the GAP model is the Hierarchical Flow Stress (HFS) model, which is shown to accurately represent the stress-strain behavior of PMMA over a broad range of strain rates and temperatures. The HFS model replicates the details of stress plateau, softening, and hardening behavior observed in glassy amorphous polymers. The theory is not restricted to isothermal conditions. Analysis is devoted to several issues associated with a non-equilibrium equation of state (EOS). Because data is insufficient for the full determination of the model's non-equilibrium EOS, several plausible conditions called the quasi-equilibrium hypothesis are put forth to complete the model. Comparisons of experimental and theoretical results over a wide-range of loading rates and conditions are reported.

1. Introduction

Consider a system consisting of large number of high molecular weight polymer molecules. On average, each polymer molecule will itself contain a large number of atoms (in actual high molecular weight polymers, average molecular weights may often reach millions). The backbone of each molecular structure is the polymer chain. For purely amorphous polymers and also for the amorphous portion of a weakly crystalline semi-crystalline polymer, these chains typically form complex networks; it is common for polymer chains to have physical entanglements, chemical cross linkages, or both (Ferry, 1980). Because of the inherent complexity at the molecular level, it is relatively easy to drive an equilibrated polymer into a long-lived non-equilibrium state by application of a simple mechanical deformation. Note that the applied deformation can be deviatoric (*i.e.*, shape changing) but also dilatometric (*i.e.* volumetric) in nature. Moreover, the ability for the polymer to relax to a new equilibrium state is a strong function of temperature and applied pressure.

It is recognized that the chain mobility is the physical quantity that governs how successful a polymer will be at relaxing back into an equilibrium configuration for the deformed system. The chain mobility is a measure of the ability of the polymer molecules to locally rearrange themselves to bring the deformed non-equilibrium structure back into a state of equilibrium (Ferry, 1980). Obviously the available thermal energy is an important factor in this process. For temperatures far above the glass transition temperature $T_g$, the chain mobility is high and a polymer is expected to relax rather quickly back into a state of equilibrium. However, below $T_g$ the chain mobility is low and the polymer system may effectively become frozen into a long-lived metastable (*i.e.* non-equilibrium) state. For temperatures below and above $T_g$ the polymer is called glassy and rubbery, respectively. The present work is primarily focused on the glassy regime but also includes temperatures slightly exceeding the glass transition.

Besides the loading rate, two experimental times are important when considering polymer relaxation. These are the total duration of the experiment and the resolution over which measurements are taken. With this in mind, we discuss two rather extreme examples.

First consider a low-rate compressive uniaxial stress experiment used to measure a polymer's stress-strain behavior. The relevant elasticity modulus is Young's modulus (or more exactly, the Young's relaxation function). Consistent with the above ideas, the observed moduli are typically strong functions of loading rate and temperature. It is useful to make the gross simplification that molecular relaxation processes can be categorized as fast, intermediate, and slow. This categorization corresponds approximately to a prototypical shear or Young's relaxation spectrum characteristic of many amorphous polymers (see Chapter 2 (Ferry, 1980) for a description of polymer behavior captured by the storage modulus). This behavior is well captured in a plot of the Young's storage modulus as a function of the load frequency for a sinusoidally loaded polymer. Figure 1 shows the different regimes for a typical entangled amorphous polymer (weakly cross-linked). High frequency processes involve only few atoms (or chain segments) in single molecules and are associated with high frequency vibrational modes, bend-and-stretch modes, and motions of side-group atoms attached to the molecular chain. At the low frequency extreme are the relaxation modes associated with the collective motion of many chain segments involving many neighboring polymer molecules and include motions like large-scale chain slippage. For an entangled polymer, relaxation times for these processes are correspondingly very long as indicated in Fig. 1. Between these extremes are the numerous intermediate rate relaxations. When a uniaxial compression experiment is done at any finite-rate of deformation, the system is immediately driven out of equilibrium. Fast relaxation processes occur so rapidly that such modes fully relax long before the first measurement is taken. However the long-lived (slow to relax) modes will continue over (or beyond) the duration of the experiment.

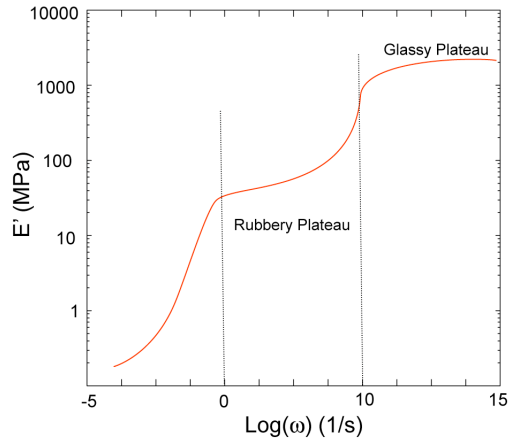

Figure 1. Youngs storage modulus as a function of loading frequency for a *generic* amorphous polymer.

The process of polymer relaxation and its effect on the mechanical behavior is encompassed in the theory of polymeric viscoelasticity. According to linear viscoelastic theory, in the absence of yield, it is the relaxation of the various modes that gives a polymer's stress-strain curve its nonlinear appearance. Because the viscoelastic response in uniaxial compression experiments is associated mainly with the deviatoric response, the relevant theory is deviatoric viscoelasticity. When the range of volumetric strain rates is limited, as in the case of a laboratory uniaxial stress-strain experiment, volumetric viscoelasticity may safely be ignored. One then invokes an "equilibrium" equation of state (EOS) to describe the volumetric behavior. A more fundamental reason for ignoring volumetric viscoelasticity is that for small volumetric changes, primarily the high frequency (for example, the vibrational modes) that are associated with fast relaxation processes are likely to be commensurate with the volumetric deformation and thus likely to be excited. As long as the experiment is long and the resolution is low, the rapid relaxation of these modes will not be observed in a relatively slow uniaxial compression experiment.

Next, consider a high velocity planar impact experiment where by the action of a shock introduced into the system the polymer will rapidly transition from an initial equilibrium state to a high pressure shocked state (Zel'dovich and Raizer, 1966; Dattelbaum and Stevens, 2008; Carter and Marsh, 1977). The duration of these experiments is typically a few microseconds (before stress relieving *release waves* interrupt the shocked state) and the experimental resolution is on the order of nanoseconds. Volumetric deformation is a dominating feature of this type of experiment. For this case we argue that volumetric relaxations from the non-equilibrium shocked state to the equilibrium shocked state, *i.e.*, volumetric viscoelasticity, are directly observable and must not be ignored. This rather innocuous claim has some rather profound scientific consequences (see for example Maugin, 1999; Prigogine, 1967; see also the discussion on the dynamic heat capacity in Wunderlich, 2007). From the practical viewpoint, this implies that the shock arrival time and particle velocity predicted from a theory based on a purely equilibrium EOS will poorly match experimentally measured values, *i.e.*, the theory will be unreliable. Because this is one of the major realizations of the present work, we summarize our findings before proceeding.

The Hugoniot is the locus of points in thermodynamic space that satisfy the Hugoniot-Rankine jump conditions (Zel'dovich and Raizer, 1966). We have found that the Hugoniot based on equilibrium measured experimental quantities does not agree with the high-rate shock measured Hugoniot. This is true for polymethylmethacrylate (PMMA) examined in detail here, but also for many other polymers that we have investigated, as mentioned below. In fact, the Hugoniot shock velocities gotten from the two different methods, for some polymers (*e.g.* PMMA) differ by nearly a factor of two when plotted in the shock velocity-particle velocity (Us-Up) plane. To a certain degree this discrepancy should be anticipated because important literature exists discussing the existence of volumetric viscoelasticity in polymers (Ferry, 1980; Sutherland 1978; Sane and Knauss, 2001; Nunziato and Schuler, 1973; Coleman *et al*., 1965; Coleman and Gurtin, 1965; Coleman and Noll, 1961). (Note that the work in these references will be discussed in Sec.

2.3.) The incorporation of volumetric viscoelasticity is needed in the present model to span volumetric rates ranging from equilibrium (essentially zero rate) to shock loading rates of $10^6$ to $10^7$ s$^{-1}$. Since many applications do not span volumetric strain rates over such a large range, volumetric viscoelasticity may be disregarded in those cases. One then resorts to invoking an "equilibrium" EOS tailored to the application's volumetric strain rate regime.

Because the present work is concerned primarily with amorphous or weakly crystalline, semi-crystalline polymers below their glass transition temperatures, the proposed model is called the Glassy Amorphous Polymer (GAP) model. In the GAP model, both the deviatoric and volumetric stresses have contributions coming from non-equilibrium behavior. That is to say, both deviatoric and volumetric viscoelasticity are part of the GAP model. A discussion of deviatoric viscoelasticity is given in Sec. 2.1 and the equilibrium EOS is given in Sec. 2.2. Upon comparing the PMMA Hugoniot calculated from the equilibrium EOS with that measured in high-rate shock experiments, the need for including a non-equilibrium EOS into the GAP model will become evident. This discussion is provided in Sec. 2.3.

An important part of the GAP model is the strength model. With increasing applied strain, many glassy polymers "flow plastically" by exhibiting a stress plateau, followed sometimes by stress softening, followed finally by stress hardening. What is in contrast to many other more brittle materials is that the relation of stress softening to damage is more ambiguous in polymers. In transparent polymers stress softening in uniaxial *compression* stress-strain behavior does not necessarily produce visible signs in the polymer such as stress whitening (Knauss, 1989; Kausch, 1978). The mechanisms of amorphous polymer flow and damage at the molecular level are probably related but in the continuum GAP model they will be treated as distinct phenomena. Thus a GAP strength model has been devised for glassy polymers to handle both softening and hardening behavior. The flow stress model acts only on deviatoric stress components - an assumption that requires future examination, but one that no available data seems to outwardly contradict. The GAP strength model, called the

Hierarchical Flow Stress (HFS) model, is discussed in Sec. 2.4. Other polymer flow stress models in the scientific literature (for example, Sarva *et al*., 2007; Richeton *et al*., 2006; Ree and Eyring, 1955) could be used if desired, but the HFS model is found to be robust and reliable. It also has interesting physical features quite different from other flow stress models.

The GAP model is developed using the compiled concepts of Secs. 2.1-2.4. The fundamental equations of the GAP model are derived from a non-equilibrium free energy. This approach is similar, but not identical to that discussed by Christensen (Christensen, 1971). A summary of the important equations in the GAP model is given in Sec. 2.5, while the details of the derivations are left to Appendices A through E. These appendices contain important discussions on non-equilibrium volumetric viscoelasticity. Due to limited experimental data, the non-equilibrium volumetric behavior must be treated approximately and two assumptions called the *quasi-equilibrium hypothesis* are introduced. These are described in Appendix B and D. Finally, it is shown that the results of this analysis can be mapped on simple deviatoric and volumetric Maxwell models, in Appendix E.

The GAP model has been applied to multiple polymers including common ones such as PMMA, PEEK, PC, Epoxy, as well as several other less common ones (Clements, Unpublished results). Only PMMA will be reviewed here for brevity and because PMMA is one of the few polymers where sufficient data exists to carry much of the required analysis of the GAP model. This is done in Sec. 3. Section 4 gives conclusions and future research directions. To focus on the crucial new ideas, a small deformation theory for the deviatoric response is used. This is different from other modeling approaches that carry out a finite deformation analysis for the deviatoric response but treat the EOS as linear elastic (two representative references are, Boyce *et al*., 1988; Richeton *et al*., 2007). Finally, polymer damage, a common component when dealing with polymer impact, is ignored here. Polymer damage will be discussed in a future publication. Mechanical rejuvenation (Struik, 1978), observed upon cyclic loading, is also ignored in the GAP model.

## 2.0 Theory

### 2.1 Devatoric Viscoelasticity

There are many books written on the topic of viscoelasticity and with few exceptions these focus primarily on deviatoric viscoelasticity. An excellent survey on viscoelasticity can be found in Ferry's book *Viscoelastic Properties of Polymers* (Ferry, 1980). At the heart of the subject is the notion that there exist characteristic relaxation times for a polymer to equilibrate once it has been driven out of equilibrium by application of mechanical or thermal agitation. Because of the practically innumerable degrees of freedom in a high molecular weight amorphous polymer, viscoelasticity theory maintains that a continuous distribution of relaxation times will represent the dynamic response of a polymer. For deviatoric viscoelasticity, the shear response is important and the shear relaxation modulus can be expressed as

$$\mu(t) = \mu^{equil} + \int_{-\infty}^{+\infty} d(\ln \tau) H(\ln \tau) e^{-t/\tau}, \tag{1}$$

where $d(\ln \tau) H(\ln \tau)$ is the distribution of relaxation times $\tau$ between $\ln \tau$ and $\ln \tau + d(\ln \tau)$. In Eq. (1), $\mu^{equil}$ is the $\tau = \infty$ contribution of $\mu(t)$, which is zero (nonzero) for a viscoelastic liquid (solid). A generalized Maxwell model (we make no distinction between Maxwell, Voigt, Kelvin, *etc.* models (Ferry, 1980)) is a discrete representation of the distribution function $H(\ln \tau)$:

$$H(\ln \tau) = \sum_{n=1}^{N} \mu^{(n)} \delta(\ln \tau - \ln \tau^{(n)}), \tag{2}$$

where $N$ is the number of deviatoric Maxwell elements in the model and $\delta$ is the Dirac delta function. The shear relaxation moduli are the coefficients $\mu^{(n)}$ and these correspond to relaxation times in the neighborhood of $\tau^{(n)}$. Substituting Eq. (2) into Eq. (1) gives the standard Prony series representation for the shear relaxation modulus

$$\mu(t) = \mu^{equil} + \sum_{n=1}^{N} \mu^{(n)} e^{-t/\tau^{(n)}} . \qquad (3)$$

Experimental methods for determining the shear relaxation moduli, relaxation times, and the temperature dependence can be found in many books on viscoelasticity (Ferry, 1980). A standard procedure that appears to work for simple amorphous polymers is to use time-temperature superposition theory to construct master curves (Williams *et al.*, 1955). Master curves are plots of the real and imaginary parts of the complex moduli (for example $\mu'(\omega)$ and $\mu''(\omega)$) over a broad range of frequencies $\omega$, at a fixed reference temperature. The real and imaginary parts are called the storage and loss moduli, respectively. By discretizing the master curve for a selected set of relaxation times, the relaxation moduli can be determined. Torsional shear or oscillatory Young's mode Dynamic Mechanical Analysis (DMA) experiments produce the storage and loss moduli versus frequency data used for master curve construction, and this has been done for many polymers in the literature (Ferry, 1980). Polymers where this procedure works are said to be thermo-rheologically simple.

The philosophy of the present work is to use DMA generated master curves only for guidance, and final adjustments to the relaxation moduli $\mu^{(n)}$ are considered justified to make the viscoelastic model agree with uniaxial stress-strain data. This approach is taken for several reasons. First, DMA is a high fidelity technique sensitive to changes in the polymer such as moisture content, synthesis and processing techniques, *etc*, and thus two DMA master curves for the "same name" polymer may differ even substantially. An example of typical variations can be found by contrasting the three PMMA DMA master curves found in (Lu *et al.*, 1997; Capodagli and Lakes, 2008). Because the GAP model neglects these effects, only approximate mechanical response can be handled. Second, variations in time-temperature construction, as well as the regimes where DMA data are gathered, again put some limitation on its use as a generic tool. Third, there is substantial discussion in the literature on if a given polymer is thermo-rheologically simple (Morland and Lee, 1960) or not. For example, polycarbonate (PC) is routinely called thermo-rheologically simple, while PMMA is not. Our goal here is not to engage in these discussions – we desire a set of relaxation moduli and relaxation times that result in a suitable fit of the experimentally measured stress-strain behavior. The next two reasons are perhaps the most important. Fourth, time-temperature master curve construction is a procedure and rigorous theoretical justification for it is not strong. Fifth, DMA is based on a perturbing strain, and in practice it is useful to extend the linear viscoelastic regime of the model to fit the experimental strain-strain behavior at somewhat higher strain values. Doing this facilitates the stress-strain fitting process in the nonlinear flow regime.

The temperature dependence of the moduli, which is a result of the master curve construction, is also justifiably modified for the above reasons to give the correct temperature dependence of the stress-strain curves. The result is a set of relaxation times, $\tau^{(n)}(T)$,

$$\tau^{(n)}(T) = a_T \tau^{(n)}(T_{ref}), \qquad (4)$$

where $a_T = a_T(T)$ is the shear shift function and $T_{ref}$ is the reference temperature. GAP uses a tabulated $a_T$ as opposed to analytic forms such as the WLF form (Williams *et al.*, 1955).

**2.2 Equilibrium EOS**

When describing polymers, equilibrium tends to be a nebulous concept, and this is also true for the volumetric behavior. Here we will define the equilibrium EOS as that which is obtained by using "slow" measurements, *i.e.*, those lasting minutes to hours. This time scale is very long compared to the deformation scenarios that the GAP model is intended. Typical slow measurements satisfying this criterion are for example the dilatometry-measured specific volume and the calorimetric-measured specific heat. Figure 2 is a plot of a family of isobars of the specific volume $v(p,T)$. A cusp in the specific volume is observed at the pressure-dependent glass transition temperature $T_g$. For many polymers, it is reasonable to treat the two regimes as having a nearly linear dependence on the temperature. We will make explicit use of this fact in our analysis described in Appendix D.

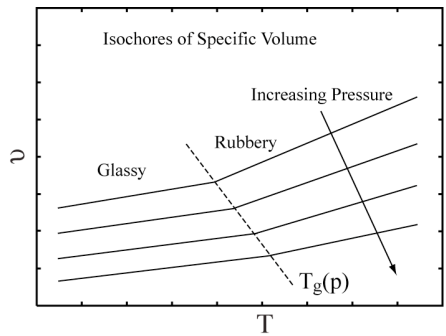

Figure 2. Equilibrium specific volume isobars as a function of temperature for a *generic* amorphous polymer, in the vicinity of the glass transition.

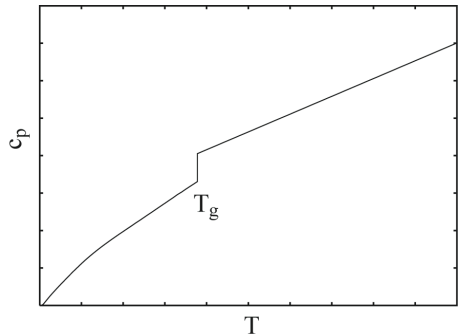

Figure 3. Equilibrium specific heat at constant pressure as a function of temperature for a *generic* amorphous polymer, in the vicinity of the glass transition.

Figure 3 shows the equilibrium specific heat at constant pressure, $c_p$, as a function of temperature for a generic amorphous polymer. As depicted, the polymer specific heat exhibits a jump discontinuity at the glass transition. Recall the introductory discussion that both $\upsilon$ and $c_p$ should be expected to have non-equilibrium behavior, *i.e.*, the behavior exhibited in Figs. 2 and 3 are the long-time limits of more general time-dependent quantities. Further discussion on the non-equilibrium behavior of the specific volume is given in Sec. 2.3, here it is mentioned that calorimetric experiments have been done to purposely apply temperature increments, usually sinusoidal, at rates fast enough to drive the polymer out of equilibrium. For these experiments the characteristic loading frequencies can be large (in excess of a 100 radians/s, for example) and are applied to investigate non-equilibrium polymeric behavior. A review has been given by Garden (2007). It is the low frequency limit of those experiments that correspond to our equilibrium specific heat.

Equilibrium specific heat (Wunderlich, 1995; Wunderlich and Pyda, 2004) and dilatometry data (Zoller and Walsh, 1995) are available for many polymers in the literature. To use that data, a simple semi-empirical expression for the Gibbs free energy $g(p,T)$ is chosen:

$$g^{(0)}(p,T) = \sum_{m=0} c_m T^m + c_L T \ln T \\ + \upsilon_0 \left[ p - R\left\{ \left(B_T^{(0)} + p\right) \ln\left(1 + p / B_T^{(0)}\right) - p \right\} \right] \qquad (5)$$

where $p$ is the pressure, and $T$ is the temperature. The reason for the superscript "(0)" will become clear in Sec. 2.5 – equilibrium is represented as the zero$^{th}$ element of a series of Maxwell elements. For the moment it is convenient to drop the superscript (0) because the entire discussion through Eq. (13) pertains only to the equilibrium state.

The dimensionless constant $R$ in Eq. (5) is equal to 0.0894. The quantities $v_0(T)$ and $B_T^{(0)}$ are related to the zero-pressure specific volume and (equilibrium) isothermal bulk modulus. The coefficients $c_M$ and $c_L$ are determined from specific heat data at ambient pressure (assumed zero), and the remaining parameters are determined from dilatometry data. Equation (5) is consistent with the Tait form (Tait, 1888) for the specific volume

$$v(p,T) = \left(\frac{\partial g}{\partial P}\right)_T = v_0(T)\left\{1 - R\ln\left[1 + \frac{p}{B}\right]\right\}, \qquad (6)$$

which has been shown to adequately model the specific volume data for many polymers. The quantities $v_0$ and $B_T$ are given the simple parametric forms

$$B_T(T) = B_0 \exp(B_1 T + B_2 T^2), \qquad (7)$$

and

$$v_0(T) = a_0 + a_1 T . \qquad (8)$$

Equation (7) is of the form used by many researchers (see for example, Quach *et al.*, 1974) in their analysis of the specific volume measurements for many polymers.

The following derivatives of $g(p,T)$ give the $p,T$-dependent equilibrium quantities required by the GAP model:

$$c_p(p,T) = -T\left(\frac{\partial^2 g}{\partial T^2}\right)_p, \qquad B_T^{-1}(p,T) = -\frac{1}{v}\left(\frac{\partial^2 g}{\partial p^2}\right)_T \qquad (9\text{-}10)$$

$$\beta(p,T) = \frac{1}{v}\left(\frac{\partial^2 g}{\partial p \partial T}\right), \qquad c_v = c_p - vB_T\beta^2 T \qquad (11\text{-}12)$$

$$\gamma = \frac{\upsilon \beta B_T}{c_\upsilon}. \tag{13}$$

In these equations $B_T$ is the equilibrium isothermal bulk modulus, $\beta$ is the equilibrium volumetric expansion coefficient, $\gamma$ is the equilibrium Grüneisen gamma coefficient and $c_\upsilon$ is the equilibrium specific heat at constant volume.

Complications arise when the glass transition or solid-solid transitions (in the case of semi-crystalline polymers) occurs in the regimes of interest. First, and to a good approximation for a reasonable range of pressures, the glass transition temperature $T_g$ depends linearly on the pressure (true for the polymers considered thus far (Clements, Unpublished results), and thus a simple parameterization will be invoked: $T_g = T_g^0 + \alpha p$, where $T_g^0$ is the ambient pressure (approximated as zero) glass transition temperature, and $\alpha$ is a constant. The means for handling the cusp in Fig. 2 and jump in Fig. 3 is to fit Eqs. (5)-(8) to the separate regimes above and below the glass transition. Likewise when a polymer (for example polytetrafluroethylene (Weir, 1954) undergoes a first-order phase transition in the crystalline portion of the polymer, a jump discontinuity occurs in the specific volume. Again, the approach is to fit the equilibrium equations to the different regimes.

**2.3 Non-equilibrium Volumetric Response**

The measured specific volume of a glassy polymer can easily become "frozen in" a long-lived non-equilibrium state (Ferry, 1980). Ferry refers to the work of Kovacs (1958) who showed that cooling a polymer at different rates, from starting temperatures slightly above the glass transition, lead to a family of specific volume curves, as depicted in Fig 4. The sharp cusp observed in the equilibrium $\upsilon$ is now rounded. Taking the departure from linearity, as approached from the rubbery polymer side, as the approximate location of the glass transition temperature, it is seen that the glass transition temperature has a rate-of-cooling dependence. Moreover, below the glass transition the departure of $v$ from its equilibrium value is a function of the rate-of-cooling in the experiment. Appendix D will make use of the observation that away from $T_g$, both regimes can still be approximately described as having linear temperature dependence.

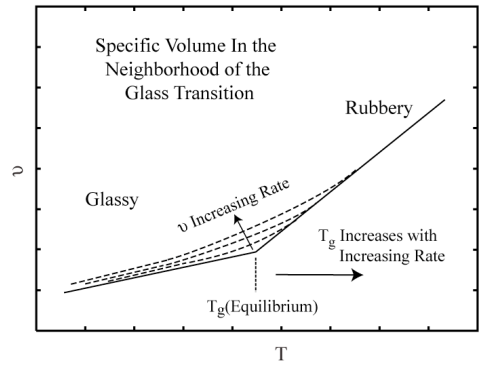

Figure 4. Isobars of the equilibrium and non-equilibrium specific volumes $v$ as a function of temperature for a generic amorphous polymer in the vicinity of the glass transition.

Next, recall the discussion in the Sec. 1 that polymer shock experiments undergoing large changes in the volumetric strain rate should be sensitive to non-equilibrium volumetric effects. To show this, a first step is to formulate the equilibrium EOS using the semi-empirical approach of Sec. 2.2. Because the equilibrium Gibbs free energy describes a thermodynamically complete EOS, the equilibrium Hugoniot may readily be calculated by determining the energy, and then determining the locus of points in thermodynamic space satisfying the Rankine-Hugoniot jump conditions (Zel'dovich and Raizer, 1966):

$$v U_S = v_0 (U_S - U_p), \tag{14}$$

$$p = p_0 + (1/v_0) U_S U_p, \tag{15}$$

$$e = e_0 + \frac{1}{2}(p + p_0)(v_0 - v).\qquad(16)$$

In these equations, $v_0$, $p_0$, and $e_0$ are initial state specific volume, pressure and specific energy, and the corresponding variables in the shocked state are $v$, $p$, and $e$. $U_S$ and $U_p$ are the shock and particle velocities, respectively. If the resulting equilibrium Hugoniot is plotted using the shock and particle velocities, comparisons can be made with the Hugoniot measured in a shock experiment (Barker and Hollenbach, 1970; Schuler, 1970). This is done in Fig. 5. The poor agreement of the experimental shock Hugoniot for PMMA (Barker and

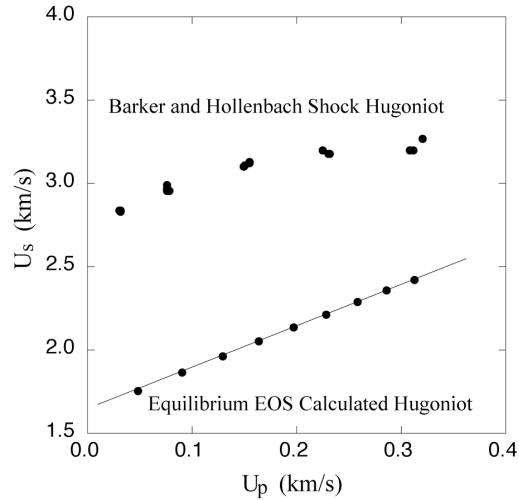

Figure 5. Shock Hugoniot from Barker and Hollenbach (1970) and that calculated from the equilibrium EOS for PMMA.

Hollenbach, 1970) and the equilibrium one is the first indication of the importance of capturing the rate dependent volumetric response. It is immediately obvious that the two Hugoniots show major discrepancies and their intercepts differ by nearly a factor of two. We speculate that this difference is due to the volumetric response of PMMA being rate dependent --- while the equilibrium EOS measurements have mechanical and thermal

"deformation" rates occurring over inverse minutes to hours, the shock measurements have a characteristic volumetric deformation rate of $10^5 - 10^7$ s$^{-1}$. We thus assert that to the equilibrium bulk modulus, bulk relaxation moduli must be added and these should increase the net bulk modulus by almost a factor of two as PMMA is shocked from equilibrium rates up to shock loading rates. A point that will become clear in Sec. 3, is that the measured shock Hugoniot in Barker and Hollenbach (1970) have undergone relaxation by several decades of strain rate, *i.e.*, they actually correspond to a rate less than $10^4$ s$^{-1}$.

Generally little data exists supporting the notion that volumetric viscoelasticity is important, however for PMMA there does exist several experiments. We now summarize some of them. Sutherland (1978) performed ultrasonic acoustic wave propagation experiments on PMMA, from which he determined the rate-dependent bulk and shear wave speeds, and their associated moduli. At rates near $10^7$ s$^{-1}$, the bulk modulus is about twice the equilibrium value (which for PMMA will turn out to be about 3.1 GPa; see Sec. 3). Next, Sane and Knauss (2001) determined the volumetric viscoelastic response for PMMA. Again they observed the bulk relaxation function to change by about a factor of two in going from very low to very high rates. Next, in the shock experiments of Barker and Hollenbach (1970) and Schuler (1970), as well as numerous recent experiments, viscoelasticity has been used to explain the rounding observed in the shock front of shocked polymers. Following that line of reasoning, Nunziato and Schuler (1973) developed a volumetric-based viscoelasticity theory and applied it to qualitatively describe the rounding of the shock front observed experimentally (Nunziato and Schuler, 1973). Finally, other experiments on polymers have been used to investigate bulk viscoelasticity (see examples given in Ch. 18 of Ferry, 1980) and find bulk moduli dependence on frequency to span a range of about a factor of two.

Thus we are led to the conclusion that not only must we deal with the equilibrium bulk modulus (Eq. (10)), but any theory that will work in regimes of both low and high volumetric strain rates should be described with a set $m = \{1,...,M\}$ of bulk relaxation times and bulk moduli $\{\bar{\tau}^{(m)}, B_T^{(m)}\}$ analogous to Eq. (3) for the shear response. The bar on the relaxation times distinguishes them from the shear relaxation times.

The question that now arises is that if the bulk response has a non-equilibrium contribution, is it necessary to measure a non-equilibrium counter part to the thermodynamic derivatives of Eqs. (11)-(13)? Our solution to this problem will be shown to be consistent with an *ansatz* for the non-equilibrium Helmholtz free energy that we judiciously chose, summarized in Sec. 2.5, and discussed in detail in the appendices.

Before leaving this section we introduce a useful measure for the volumetric "departure" from equilibrium. Disregarding the differences of adiabatic and isothermal bulk moduli, the $U_p = 0$ intercept of Fig. 5 is a bulk sound speed. The difference between the equilibrium bulk modulus $B^{(0)}$ and the relevant modulus at shock-rates, *i.e.*, so the called instantaneous modulus (Nunziato and Schuler, 1973), $B^{(0)} + \sum B^{(m)}$, is simply $\sum B^{(m)}$. A useful parameter, $\xi(M')$, which is a generalization to any volumetric strain rate, and tends to lie in the range $0 \leq \xi \leq 1$ is

$$\sum_{m=1}^{M'} B^{(m)} / B^{(0)} \equiv \xi(M'). \tag{17}$$

$M' \leq M$ is defined as the largest *active* value of $m$. Letting $\dot{\varepsilon}_v$ denote the current volumetric strain rate, active elements, $m$, satisfy $\dot{\varepsilon}_v \bar{\tau}^{(m)} \gg 1$ and are included in the sum in Eq. (17). Inactive elements satisfy $\dot{\varepsilon}_v \bar{\tau}^{(m)} \ll 1$. Thus $m = M'$ will approximately satisfy $\dot{\varepsilon}_v \bar{\tau}^{(m)} \approx 1$. Parenthetically, because the relaxation times are typically expressed in decade increments, this gross description is sensible. Note that $\xi(M')$ will enter the discussion of the effective Grüneisen Gamma coefficient in Sec. 2.5 and the appendices.

**2.4 GAP Hierarchical Flow Stress (HFS) Model**

In uniaxial stress experiments many glassy polymers exhibit strong deviations from linear viscoelastic behavior as the strain exceeds a few percent. It is common in the glassy polymer literature to refer to the yield plateau, softening, and hardening behavior observed experimentally in uniaxial stress-strain curves as plastic flow. Interestingly, stress softening is usually associated with damage and localization (for example, necking). In glassy polymers, however, stress softening under standard uniaxial compression stress conditions is ambiguously associated with macroscopic damage (or localization) in regimes of small strains. To expound on this, stress whitening observed in deforming transparent polymers is an indication of strong molecular alignment or compaction (thus changing its translucency). When a tensile strain component is present, stress whitening is a well-known beginning stage of craze formation, a precursor to damage. Polycarbonate, a transparent polymer, remains visibly transparent as the stress yields and then softens under a controlled compressive straining. If damage is occurring in polycarbonate, it is occurring at the molecular level and produces no visible signs at the macroscopic level. This example demonstrates that the relation of stress softening to damage is ambiguous in polymers. Because of this, stress softening will be incorporated into the GAP theory *via* the plasticity model.

In the GAP model only deviatoric plasticity is considered and each shear Maxwell element, $n \in \{0,...,N\}$, is assigned a deviatoric plastic strain. The fact that plastic strain is associated with each of the $N$ elements is different from other plasticity models (see for example, Sarva *et al.*, 2007; Richeton *et al.*, 2006; Ree and Eyring, 1955; Bauwens-Crowet, 1973; Robertson, 1966; Eyring, 1936). In those models, the often observed linearity of the yield plateau with the logarithm of strain rate has motivated the view that yield is associated with only a single (Eyring, 1936) (and sometimes a double (Bauwens-Crowet, 1973)) activated process. While a single or double activated process may suffice in limited regimes, no assertion will be made here that that is true in general. Fitting the flow stress model to stress-strain experiments will in principle determine how weak or strong any particular process will contribute to the overall yield process.

Plasticity in the GAP model is introduced through a modified Prandtl-Reuss (see for example, Hill, 1950) viscoplastic flow law

$$de_{ij}^{P(n)} = \Lambda_P^{(n)} s_{ij}^{(n)} dt , \qquad (18)$$

where the flow rule function has a form resembling that used by Bodner-Partom (1975) and Frank and Brockman (2001):

$$\Lambda_P^{(n)}(t) = \frac{D_0}{\sqrt{J_2^{(n)}(t)}} \left( \frac{\dot{\varepsilon}_{eff}(t)}{\dot{\varepsilon}_{eff}^0} \right)^{n_z} \left( \frac{3 J_2^{(n)}(t)}{\left[ f(t) Z^{(n)}(t) \right]^2} \right)^{n_p} , \qquad (19)$$

but with several important modifications. First, as mentioned above, each Maxwell element $n$ will flow plastically according to the deviator, $s_{ij}^{(n)}$, of that element. For a element deviator $s_{ij}^{(n)}$ to contribute significantly to the total deviator stress, $s_{ij}$, it must be active and as a rule of thumb, the applied strain rate must then satisfy $\tau^{(n)} \dot{\varepsilon} \gg 0$, where $\dot{\varepsilon}$ is a measure of the applied strain rate. In the sense that the flow rule is a function of $n$, the GAP flow theory is similar to that proposed by Frank and Brockman (2001).

In Eq. (19), $Z^{(n)}(t)$ is a load-history-dependent variable related to microstructural arrangements occurring during straining. The primary purpose of $Z^{(n)}(t)$ is to capture stress yielding (plateau), softening, and hardening structure observed as a function of strain. $Z^{(n)}(t)$ is also assigned a Maxwell-element dependence through the index $n$. The material parameter $D_0$ is the limiting shear strain rate. The factor $2J_2^{(n)} \equiv s_{ij}^{(n)} s_{ij}^{(n)}$ is the second invariant of the deviatoric stress for element $n$. For the present purpose, the material parameters $n_z$ and $n_p$ are rate sensitivity parameters and are determined by fitting uniaxial stress stress-strain experiments. The effective strain rate (Hill, 1950) in Eq. (19) is given by

$$\dot{\varepsilon}_{eff} = \sqrt{\frac{2}{3} \dot{e}_{ij} \dot{e}_{ij}} \ . \tag{20}$$

Letting $\bar{W}_P^{(n)}$ denote the plastic work per unit volume for element $n$, the corresponding plastic work increment is given by

$$d\bar{W}_P^{(n)} = s_{ij}^{(n)} de_{ij}^{P(n)} \ . \tag{21}$$

The bar over the work distinguishes it from the work per unit mass used in the thermodynamic part of the GAP theory. $Z^{(n)}(t)$ is comprised of a term $Z_0^{(n)}(t)$ plus an additional work hardening contribution proportional to $\bar{W}_P^{(n)}$:

$$Z^{(n)}(t) = Z_0^{(n)}(t) + \alpha_W \bar{W}_P^{(n)}(t) \ , \tag{22}$$

where $\alpha_W$ is a unitless material parameter. Equation (22) is again similar to that used by Frank and Brockman (2001). In the GAP flow model, the accumulated plastic work is still used as a scalar measure of the plastic strain achieved during deformation.

The second primary difference of the GAP plastic flow theory and that of Bodner-Partom (1975) and Frank and Brockman (2001) is that the latter theories consider yielding and subsequent work hardening, whereas the present theory must also include stress softening. To that end, the $Z_0^{(n)}(t)$ evolve according to a hierarchical set of ordinary differential equations (ODEs)

$$dZ_j^{(n)}(t) = m_j \left(1 - \frac{Z_j^{(n)}(t)}{Z_{j+1}^{(n)}(t)}\right) d\overline{W}_P^{(n)}(t), \qquad j = 0,...,J \qquad (23)$$

with initial conditions

$$Z_j^{(n)}(t=0) = K_j Z_0 \left(\frac{\mu^{(n)}}{\sum_{k=0}^{} \mu^{(k)}}\right) \equiv K_j Z_0 \Gamma^{(n)}(\mu^{(n)}) \qquad n = 0,...,N, \quad j = 0,...,J+1. \qquad (24)$$

For the $n^{th}$ element, $Z_j^{(n)}(t=0)$ scales with the distribution of shear relaxation moduli, $\Gamma^{(n)}$, for that element. This scaling is in accord with molecular co-operativity theory as mentioned by Frank and Brockman (2001). $Z_o$ is a material constant and has dimensions of stress. The unitless constants $K_j$ vary slightly about unity and are chosen such that plasticity onset, yield, softening, and final hardening are captured. Five values of $j$ typically suffice, and thus the hierarchical series of evolution equations is truncated at $J=4$, with $Z_5^{(n)}$ being specified as a material constant. $K_j$ and $m_j$ are regarded as material fitting parameters. Because this hierarchical approach is novel (from which the name Hierarchical Flow Stress (HFS) is given to the model), some discussion is prudent.

It will suffice to discuss the behavior of Eqs. (23) and (24) for a single $n \in \{0,...,N\}$. Moreover for a given $n$, the distribution of shear relaxation moduli $\Gamma^{(n)}$ contributes only a constant scale factor to $Z_0$. Also, the effects of including work hardening are clear in Eq.

(22), and thus $\alpha_W = 0$ is taken here to simplify the present discussion. In actual calculations, $\alpha_W$ is not zero. Finally, because $K_j Z_o \Gamma^{(n)}$ appears only as a product in Eq. (24), only the product is important for each $j$. The other important material parameter is $m_j$.

For simple loads (say constant strain rate uniaxial stress loading) the solution of Eq. (23), subject to the initial conditions (Eq. (24)), is relatively easy to describe. First, note that the solution of *each* ODE (*i.e.*, each $j$) will begin at $Z_j^{(n)}(t=0) = K_j Z_0 \Gamma^{(n)}$. Next, it is not difficult to demonstrate that for sufficiently large values of $\bar{W}_P^{(n)}$, *every* $Z_j^{(n)}(t)$ of the hierarchy individually converges to $K_5 Z_0 \Gamma^{(n)}$. The rate of convergence from the initial (*i.e.*, $K_j Z_0 \Gamma^{(n)}$) to final value (*i.e.*, $K_5 Z_0 \Gamma^{(n)}$) is controlled by $m_j$ which multiplies $\bar{W}_P^{(n)}$ in Eq. (23). Large values of $m_j$ cause a quick transition to $K_5 Z_0 \Gamma^{(n)}$ (*i.e.*, the transition will occur at small strains) while small $m_j$ will exhibit a slow transition (*i.e.*, the transition will occur at large strains). Because the solution of the $j^{th}$ state variable $Z_j^{(n)}(t)$ depends on the history of the $j+1^{th}$ state variable $Z_{j+1}^{(n)}(t)$ in Eq. (23) the transition will generally be non-monotonic. The exception is $Z_4^{(n)}(t)$ because $Z_5^{(n)}$ is a constant. By judicious choices of the pairs $(K_j, m_j)$ it is a relatively easy matter to adjust the flow stress (with $K_j$) and strain location (with $m_j$) to agree with the observed yield, softening, and hardening behavior.

The physical motivation for the hierarchical approach can be explained. According to the HFS model, at any given point in the deviatoric deformation all plastic flow processes are present *via* the hierarchy but not all of those processes are weighted equally. For example, at small strains (little accumulated plastic work) the molecular aligning or compaction processes important at large strains enter through the hierarchy but have very little influence

(weight) and only begin to dominate at large strains. An advantage of the HFS model is that by using a discrete representation (a set of five total $j$) the hierarchical approach allows us to continuously interpolate between the innumerable molecular level processes much the same as the discrete Maxwell model representation does for the innumerable molecular relaxation processes responsible for viscoelasticity.

Before leaving this section, a few remaining points require discussion. The first point has to do with redundancy in the HFS model. It can be shown for a large number of polymers (Clements, Unpublished results) that if the total number of Maxwell elements $N$ used in the analysis is large, suitable agreement with experimental stress-strain curves over wide ranges of strain rates and temperatures can be achieved by setting $f(t) = 1$ and $n_z = 0$. However, computational efficiency demands using a small number of Maxwell elements (say less than a dozen), thus two factors are added into the HFS model to allow for agreeable comparisons to measured stress-strain curves. First, a scale factor $f(t)$, multiplying $Z^{(n)}(t)$ is added. This enables further temperature and strain rate sensitivity:

$$f = 1 + f_T \log(a_T) + f_\varepsilon \log(\dot{\varepsilon}_{eff} / \dot{\varepsilon}_{eff}^0). \tag{25}$$

The form of Eq. (25) recognizes that the relevant temperature and strain rate for polymers is derived through the shift function $\log(a_T)$ and the $\log(\dot{\varepsilon}_{eff})$, respectively. $f_T$ and $f_\varepsilon$ are material constants. Secondly, additional rate sensitivity is included by adding the strain-rate dependent prefactor to Eq. (19).

There is also potential redundancy in the HFS theory by including the work hardening term in Eq. (22) together with work hardening *via* the large $j$ state variables (*e.g.* $Z_5^{(n)}$ in the hierarchy). The additional work hardening sensitivity gained by including Eq. (22) is done for convenience as it affords a little more flexibility in the model.

As a final point, the HFS model has an implicit pressure dependence coming through the volumetric compression dependence of the shear relaxation moduli (see Eq. (31) in Sec. 2.5). This pressure dependence enters the numerator of Eq. (24), but not in the denominator (which for a given polymer and Maxwell representation is a fixed number). Equivalently then, this has the same effect on the HFS model as giving a pressure dependence to $Z_0$.

**2.5 Summary of the GAP model**

In this section the GAP model is summarized and detailed derivations are left to Appendices A through E. The equations are derived from a non-equilibrium Helmholtz free energy, and are expected to obey thermodynamic consistency up to several articulated caveats. The pressure increment in the GAP model has contributions coming from the equilibrium EOS, $dp^{(0)}$, plus non-equilibrium contributions

$$dp = dp^{(0)} + \sum_{m=1}^{M} dp^{(m)}, \qquad (26)$$

where the $m = 1,...,M$ Maxwell incremental pressures are

$$dp^{(m)} = B_T^{(m)} d\varepsilon_v + \beta^{(0)} B_T^{(m)} dT - \frac{p^{(m)}}{\tau^{(m)}} dt, \quad m = 1,...,M. \qquad (27)$$

A similar expression exists for the equilibrium pressure increment $dp^{(0)}$, gotten by substituting $m = 0$ into Eq. (27), but with the omission of the relaxation term.

In Eq. (27), $B_T^{(m)}$ is the $m^{\text{th}}$ isothermal bulk modulus, $\bar{\tau}^{(m)}$ is the corresponding bulk relaxation time, $d\varepsilon_v = -d\ln v$ is the volumetric strain increment chosen to be positive in compression (see Appendix A), and $v$ is the specific volume. Finally, $dT$ and $dt$ are the temperature and time increments, respectively. Discussed in detail in Appendix B and D, the two conditions comprising the *quasi-equilibrium hypothesis* inherent in this work are:

1. The $m = 0,..., M$ elements are at a common temperature $T$.
2. All elements of the non-equilibrium volume expansion coefficient, $\beta^{(m)}$, are equal to the equilibrium value $\beta^{(0)}$.

The equilibrium isothermal bulk modulus $B_T^{(0)}$ is derived (see Sec. 2.2) from the equilibrium EOS, and has volume and temperature dependence determined accordingly. In contrast, the non-equilibrium elements of the isothermal bulk relaxation moduli, $B_T^{(m)}$, $m = 1,..., M$, are not derived from more fundamental thermodynamic quantities, but rather by construction they are given volumetric strain dependence through a simple polynomial parameterization:

$$B_T^{(m)}(t) = B_T^{(m)}\left(1 + \chi_1 \varepsilon_v(t) + \chi_2 \varepsilon_v^2(t)\right) \quad m \geq 1 , \tag{28}$$

where $\chi_1$ and $\chi_2$ are material parameters, and are zero in volumetric expansion.

In principle, the temperature dependence of the volumetric response can be deduced in the same manor as is done for the shear relaxation moduli described below, *i.e.*, through the temperature dependence of the volumetric relaxation times $\bar{\tau}^{(m)}$. A usual procedure to determine the temperature dependence would be to construct a time-temperature master curve (Ferry, 1980). Unfortunately, this procedure is seldom done for the volumetric response, and master curve data for the volumteric response does not typically exist for glassy polymers. PMMA is one of the few exceptions (Sane and Knauss, 2001). The second most common approach, while less direct, is to deduce the temperature dependence by comparing to experiments (for example stress strain experiments) done over a broad range of temperature. Again, for the deviatoric response there is often sufficient experimental data available to carry out such an analysis, while for the volumetric response there is insufficient available data that couples sufficiently well to the bulk response to unambiguously determine the temperature dependence of the $B_T^{(m)}$.

The deviatoric stress increment can also be split into equilibrium and non-equilibrium terms according to

$$ds_{ij} = ds_{ij}^{(0)} + \sum_{n=1}^{N} ds_{ij}^{(n)} . \qquad (29)$$

where the $n = 1,...,N$ incremental deviatoric stresses are given by

$$ds_{ij}^{(n)} = 2\mu^{(n)}\left(de_{ij} - de_{ij}^{P(n)}\right) - \frac{s_{ij}^{(n)}}{\tau^{(n)}} dt, \quad n = 1,...,N . \qquad (30)$$

In Eq. (30), $\mu^{(n)}$ are the shear relaxation moduli, with associated shear relaxation times, $\tau^{(n)}$, $de_{ij}$ are the components of the deviatoric strain increment tensor, and $de_{ij}^{P(n)}$ are the deviatoric plastic strain increments of the $n^{th}$ Maxwell shear element. The equilibrium contribution, $n = 0$, of the deviatoric stress increment $ds_{ij}^{(0)}$ is given by Eq. (30) but with the omission of the relaxation term. Unlike the volumetric case, the equilibrium shear modulus, $\mu^{(0)}$, is not obtained from an equilibrium free energy. Rather it is assumed to be temperature independent and its volumetric dependence is given by the simple parameterization

$$\mu^{(n)}(\varepsilon_\upsilon) = \mu^{(n)}\left[1 + \chi_3 \varepsilon_\upsilon\right], \quad n = 0,...,N \quad \varepsilon_\upsilon > 0, \tag{31}$$

where $\chi_3$ is a constant material parameter. The pressure is known to influence the shear response of many polymers (Sauer, 1977; Sauer *et al.*, 1970; Pae and Bhateja, 1975), and for low pressures a linear $\varepsilon_\upsilon$ dependence on $\mu^{(n)}$ should suffice.

The temperature increment in the GAP model is derived in Appendix B, and is

$$dT = \frac{1}{c_\upsilon^{eff}} \sum_{m=0}^{M} T ds^{(m)} - \rho T \gamma_{eff} d\upsilon - \frac{T\upsilon}{c_\upsilon^{eff}} \sum_{m=0}^{M} \beta^{(0)} \frac{p^{(m)}}{\overline{\tau}^{(m)}} dt, \tag{32}$$

where an effective Grüneisen gamma coefficient $\gamma_{eff}$ and specific heat $c_\upsilon^{eff}$ have been introduced:

$$\gamma_{eff} = \frac{1}{c_\upsilon^{eff}} \sum_{m=0}^{M} \gamma^{(m)} c_\upsilon^{(m)}, \tag{33}$$

where

$$c_\upsilon^{eff} = \sum_{m=0}^{M} c_\upsilon^{(m)}. \tag{34}$$

It is also argued in the Appendix D that a suitable approximation is to take

$$\gamma_{eff} \approx \gamma^{(0)}(1 + \xi(M')) \quad \text{and} \quad c_\upsilon^{eff} \approx c_\upsilon^{(0)} \tag{35}$$

where , $\gamma^{(0)}$ and $c_v^{(0)}$ are the equilibrium Grüneisen coefficient and specific heat, respectively. The quantity $\xi(M')$ can be estimated once the high rate experimental Hugoniot obtained in a shock experiment (Sec. 2.3) is known, along with the equilibrium-calculated one obtained from the equilibrium measured quantities, as described in Sec. 2.2, *i.e.*,

$$\sum_{m=1}^{M'} B_T^{(m)} \equiv \xi(M')B_T^{(0)} . \tag{36}$$

As derived in Appendix (B), $Tds$ in Eq. (32) is given by

$$\rho Tds = \sum_{m=1}^{M} \frac{p^{(m)}p^{(m)}}{B_T^{(m)}\overline{\tau}^{(m)}} + \sum_{n=1}^{N} \frac{s_{ij}^{(n)}s_{ij}^{(n)}}{2\mu^{(n)}\tau^{(n)}} + \sum_{n=0}^{N} s_{ij}^{(n)}de_{ij}^{P(n)} . \tag{37}$$

The right-most term involves the plastic work increment discussed in Sec. 2.4.

Before leaving this section, several words on the numerical solution scheme presently invoked are in order. Because the same scheme is used for both the deviatoric and the volumetric equations, only the deviatoric solution is described. The intent is to probe high rate dynamic deformation problems, and thus an explicit time integration solution scheme is used, *i.e.*, the time increment is assumed to be very small. An approximate, but stable solution scheme has been derived to solve Eq. (30). Omitting the details, Eq. (30) can be written in a sequential form as

$$s_{ij}^{(n)}(t) \approx a^{(n)}s_{ij}^{(n)}(t-\Delta t) + 2\mu^{(n)}(t)b^{(n)}(t)de_{ij} \tag{38}$$

where $b^{(n)}$ gives weight to only those elements that are active for a given strain rate:

$$b^{(n)} = \frac{1-e^{-\Delta\phi^{(n)}}}{\Delta\phi^{(n)}} \equiv \frac{1-a^{(n)}}{\Delta\phi^{(n)}}. \tag{39}$$

Note that when significant material rotation is present Eq. (38) will be modified to include the Jaumann-Zaremba rate as described for example in Holzapfel (2007).

In Eq. (39),

$$\Delta\phi^{(n)} = \phi^{(n)}(t) - \phi^{(n)}(t-\Delta t) = \int_{t-\Delta t}^{t} dt' \left[ \frac{1}{\tau^{(n)}(t')} + 2\mu^{(n)}\Lambda^{(n)}(t')\chi(t') \right], \tag{40}$$

where the last term in the brackets is related to the plastic strain increment *via* Eq. (18). In the solutions to the GAP model presented the integral in Eq. (40) is approximated as a simple trapezoid integration. Of coarse as $dt$ gets smaller and smaller the approximate solutions become more precise. Clearly, more sophisticated integrations can be invoked as desired.

## 3. Results and Discussion

The analysis outlined in the previous sections is applicable to many amorphous glassy polymers (Clements, Unpublished results). In this section, parameters, analysis and results specific to PMMA will be given. For wave propagation problems the density is needed and for this we use 1185 kg/m³. Aside from the equilbrium EOS calculation, all simulations were done by implementing the GAP model into an Abaqus VUMAT (Abaqus, 2009). Dynamic simulations, including the stress-strain calculations, were done assuming 2-D axial symmetry.

A tabulated specific heat for PMMA is given in the ATHAS polymer database (the ATHAS database can be found at http:/web.utk.edu/~athas) and specific volume isobars between 0 and 700 MPa have been measured by Theobald, Pechhold and Stoll (Theobald *et al.,* 2001). This information is sufficient to determine the equilibrium Gibbs free energy parameters in Eqs. (5)-(8) and these are listed in Table 1. Plots of the corresponding equilibrium specific Gibbs free energy (Fig. 6), specific volume (Fig. 7), equilibrium $c_p$ (Fig. 8), and equilibrium isothermal bulk modulus (Fig. 9) are shown. Each plot shows a family of isobars between 0 and 700 MPa. The pressure dependent glass transition temperature is described by $T_g = T_g^0 + \alpha p$, where $T_g^0 = 100°C$ and $\alpha = 1.6\text{E-}07$ Pa. In Eqs. (28) and (31), $\chi_1$, $\chi_2$, and $\chi_3$ are set equal to 7.0, 24.0, and 0.6 respectively. These parameters give a volumetric strain dependence of the bulk and shear moduli. To determine $\chi_3$ we used the PMMA shear moduli data of Rabinowitz, Ward, and Parry (Rabinowitz *et al.,* 1970). Values for $\chi_1$ and $\chi_2$ were chosen to make the PMMA properly shock-up in the shock analysis described below.

Table 1. Equilibrium EOS parameters for PMMA from fitting experimental $c_P$ and $\upsilon$ data.

|  | $c_0$ (J / Kg) | $c_1$ (J / (Kg K)) | $c_2$ (J / (Kg K²)) | $c_3$ J / (Kg K³) | $c_4$ J / (Kg K⁴) | $c_L$ J / (Kg K) |
|---|---|---|---|---|---|---|
| $T < T_g$ | 6.8E+04 | 11.0 | -3.0 | 9.5E-04 | -2.75E-07 | 0.0 |
| $T > T_g$ | -8.7E+04 | 3.2E+03 | -2.75 | 7.917E-04 | -1.917E-07 | -480.0 |
|  | $B_0$ (Pa) | $B_1$ (°C⁻¹) | $B_2$ (°C⁻²) | $a_0$ (m³ / Kg) | $a_1$ (m³ / (Kg °C)) |  |
| $T < T_g$ | 2.866E+08 | -1.95E-03 | 1.E-06 | 8.380E-04 | 1.8316E-07 |  |
| $T > T_g$ | 3.035E+08 | -4.427E-03 | 1.E-06 | 8.035E-04 | 5.2016E-07 |  |

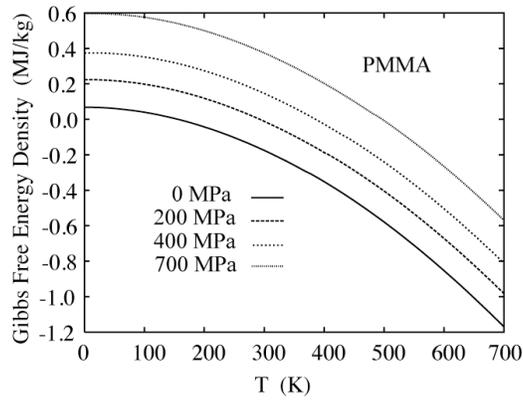

Figure 6. Equilibrium Gibbs free energy density for PMMA, calculated from Eq. 5.

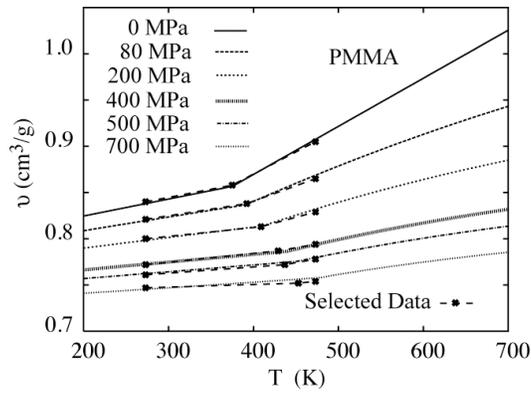

Figure 7. PMMA equilibrium $\upsilon$ (Eq. 6). Selected points are from Theobald *et al*. (2001).

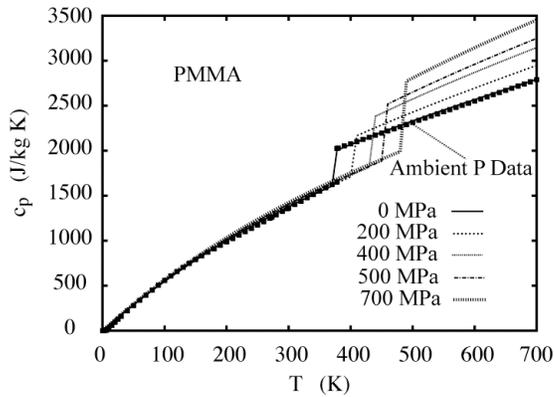

Figure 8. Equilibrium specific heat for PMMA (Eq. 9). The ambient pressure curve is from the ATHAS database. The jump discontinuity is at the glass transition.

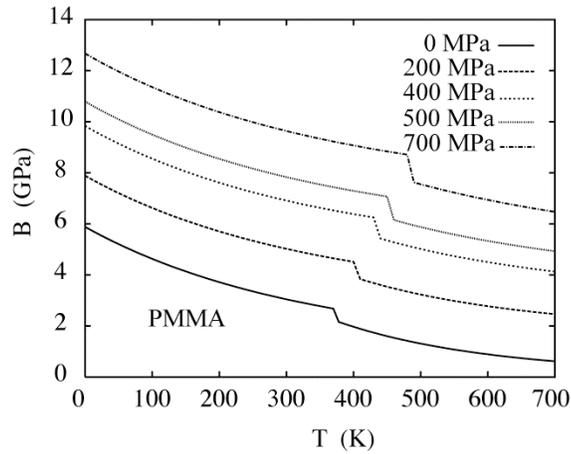

Figure 9. Equilibrium isothermal bulk modulus for PMMA from the Eq. (10). Note that the ambient pressure, room temperature value is about 3.1 GPa.

Uniaxial compression stress-strain data have been measured for PMMA (Richeton *et al.*, 2006) for sufficient temperature and strain rate variation to satisfactorily determine the parameters for the viscoelastic and plastic flow model of Sec. 2.1 and 2.4, respectively. The resulting shear relaxation times and moduli that give acceptable agreement with the small-strain stress-strain data are listed in Table 2. A plot of corresponding shear shift function is shown in Fig. 10.

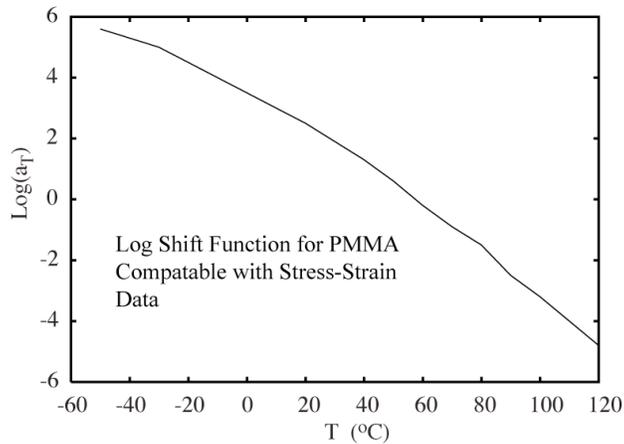

Figure 10. PMMA logarithm of the shear shift function, $a_T$.

Table 2. Shear relaxation times and moduli for PMMA.

| $n$ | $\tau^{(n)}$ (s) | $\mu^{(n)}$ (GPa) |
|---|---|---|
| 0 |  | 0.20 |
| 1 | 1.E+04 | 0.08 |
| 2 | 1.E+03 | 0.06 |
| 3 | 1.E+02 | 0.07 |
| 4 | 1.E+01 | 0.08 |
| 5 | 1.E+00 | 0.09 |
| 6 | 1.E-01 | 0.13 |
| 7 | 1.E-02 | 0.16 |
| 8 | 1.E-03 | 0.20 |
| 9 | 1.E-04 | 0.26 |
| 10 | 1.E-05 | 0.30 |
| 11 | 1.E-06 | 0.34 |
| 12 | 1.E-07 | 0.38 |

The HFS model is parameterized from the high strain regions of the stress-strain curves of (Richeton *et al.*, 2006). Specifying the effective strain rate to be $\dot{\varepsilon}^0_{eff} = 1.\text{e-}04$ s$^{-1}$, the material parameters of Eqs. 19 and 22-25 required to fit the stress-strain curves of Figs. 11 and 12 are listed in Table 3. The high rate data (900, 2300, and 4300 s$^{-1}$) in Fig. 13 is taken from Split Hopkinson pressure bar measurements (Richeton *et al.*, 2006) and represent nominal strain rates and averaged stresses. For rates of 10 s$^{-1}$ and above, it is assumed that isothermal conditions are not maintained in the experiments, *i.e.*, PMMA heats according to Eqs. 32 and 37, of Sec. 2.5. For the rates below 10 s$^{-1}$, isothermal conditions were applied. The softening in the stress strain curves at rates of 10 s$^{-1}$ and above is a direct consequence of the polymer heating, referred to in the literature as adiabatic heating and discussed by other modeling efforts as well (see for example Richeton *et al.*, 2006).

Table 3. PMMA HFS model parameters. Unless specified, parameters are dimensionless.

| $K_0$ | 1.0 | $m_0$ | 540.0 | $n_z$ | 1.0 |
|---|---|---|---|---|---|
| $K_1$ | 1.7 | $m_1$ | 70.2 | $n_p$ | 10.0 |
| $K_2$ | 1.6 | $m_2$ | 16.2 | $D_0 (s^{-1})$ | 1.e+04 |
| $K_3$ | 1.3 | $m_3$ | 8.1 | $f_T$ | 0.05 |
| $K_4$ | 0.8 | $m_4$ | 6.7 | $f_\varepsilon$ | 1.e-04 |
| $K_5$ | 0.9 | $\alpha_W$ | 1.1 | $Z_0 (Pa)$ | 5.5E+08 |

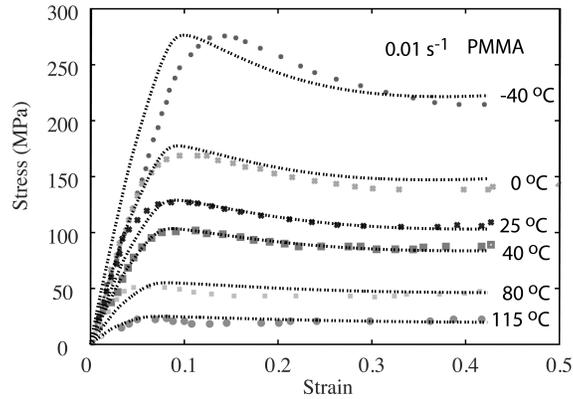

Figure 11. Temperature dependence of the compressive stress-strain behavior for PMMA at a fixed strain rate of 0.01 s$^{-1}$. Points are selected data from (Richeton *et al.*, 2006).

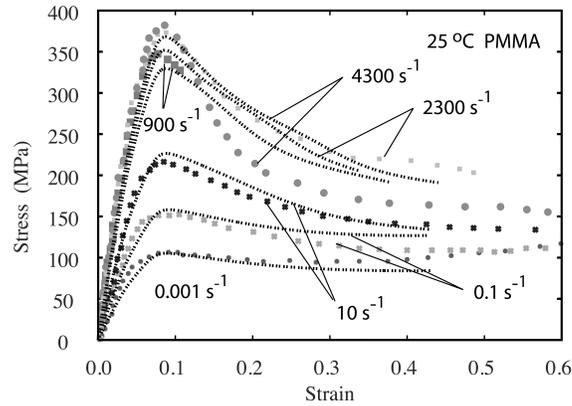

Figure 12. Strain-rate dependence of the compressive stress-strain behavior for PMMA at 25°C. Points are selected data from (Richeton *et al.*, 2006).

We now turn to the non-equilibrium volumetric part of the GAP model. For many polymers limited data exists on the non-equilibrium volumetric response and this presents a problem for the model developer. The situation is better for PMMA (see the discussion in Sec. 2.3) and here we will use the dynamic bulk compliance data of Sane and Knauss (2001). We approximate their bulk storage compliance as the inverse of the bulk storage modulus. This is a fair approximation when the bulk loss compliance is small - the case for PMMA (Sane and Knauss, 2001). The PMMA master curve of Sane and Knauss (2001) used a reference temperature of 105°C. With a linear fit to their bulk shift data, a volumetric shift factor of 7.65 shifts the data back to room temperature. The data of Ref. [10], inverted to give $B'(\omega)$, with this shift is displayed in Fig. 13 along with two theoretical isothermal bulk storage moduli $B'(\omega)$ calculated from (Ferry, 1980):

$$B'(\omega) = B_T^{(0)} + \sum_{m=1}^{M} B_T^{(m)} \frac{\left(\omega \bar{\tau}^{(m)}\right)^2}{1+\left(\omega \bar{\tau}^{(m)}\right)^2} . \qquad (41)$$

The two $B'(\omega)$ correspond to a 6+1 ("+1" being the equilibrium element) and 12+1 set of bulk relaxation times and relaxation moduli. Table 4 lists the values for both the 6+1 and

12+1 model. Both the 6 and 12 non-equilibrium elements add up to the same value of 3.9 GPa. Recall that the equilibrium $B_T^{(0)}$ is derived from the equilibrium Gibbs free energy (and is a function of temperature and pressure) from Sec. 2.2. The staircase structure at low frequencies observable in Fig. 13 arises because of the coarseness of relaxation times used in that regime. The 12+1 fit is appreciably better in that regard, however the 12+1 fit produces no noteworthy change in stress strain curves or fits to shock profiles discussed below. To minimize computation expense as well as computer memory allocation, only the 6+1 fit is used in the calculations and discussed hereafter. A simple linear fit to the bulk shift function of Sane and Knauss (2001) suffices: $\mathrm{Log}(\bar{a}_T) = 1.8 - 0.09\,T(^\circ C)$, and is adjusted to make the reference temperature 20°C. As described below, the 6+1 element model is found to be almost exactly what is needed to improve the Hugoniot calculation, and to within an acceptable tolerance, fits the data of Sane and Knauss (2001). It should not be understated that this observation provides strong validation for the statement that volumetric viscoelasticity is the required missing physics in a purely equilibrium volumetric analysis.

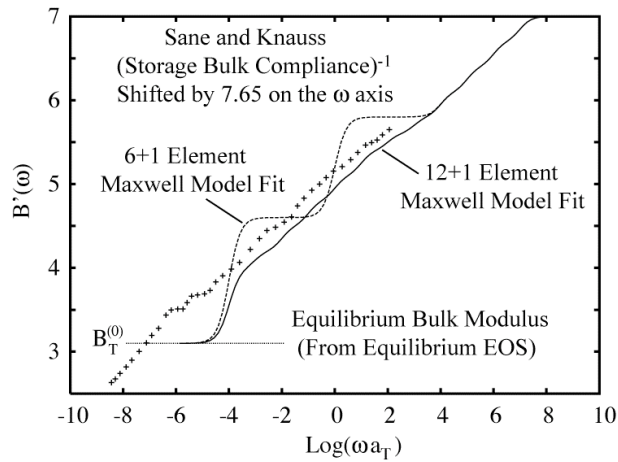

Figure 13. Bulk storage modulus for PMMA at 20°C for 6+1 and 12+1 generalized Maxwell models for the volumetric response. Points are selected data from Sane and Knauss (2001).

Table 4. Bulk relaxation times and moduli for PMMA for a 6+1 and 12+1 generalized Maxwell model for the volumetric response.

| $m$ | $\bar{\tau}^{(m)}$ (s) | $B_T^{(m)}$ (GPa) | $m$ | $\bar{\tau}^{(m)}$ (s) | $B_T^{(m)}$ (GPa) | $m$ | $\bar{\tau}^{(m)}$ (s) | $B_T^{(m)}$ (GPa) |
|---|---|---|---|---|---|---|---|---|
| 1 | 1.E+04 | 1.5 | 1 | 1.E+04 | 0.9 | 7 | 1.E-02 | 0.2 |
| 2 | 1.E+00 | 1.2 | 2 | 1.E+03 | 0.2 | 8 | 1.E-03 | 0.2 |
| 3 | 1.E-04 | 0.3 | 3 | 1.E+02 | 0.3 | 9 | 1.E-04 | 0.3 |
| 4 | 1.E-05 | 0.3 | 4 | 1.E+01 | 0.3 | 10 | 1.E-05 | 0.3 |
| 5 | 1.E-06 | 0.3 | 5 | 1.E+00 | 0.3 | 11 | 1.E-06 | 0.3 |
| 6 | 1.E-07 | 0.3 | 6 | 1.E-01 | 0.3 | 12 | 1.E-07 | 0.3 |

We now return to the analysis of a shock propagating in PMMA. (Note that a brief slightly modified version of this work can be found in (Clements, 2009.) Barker and Hollenbach (1970) carried out symmetric impact (a PMMA flyer plate impacting a PMMA target) experiments and the particle velocity profiles are shown in Fig. 14. The flyer plate thickness is approximately 6.35 mm and the velocity profile is measured about the same distance into the target. GAP model velocity profiles for multiple impact speeds (listed in the figure) are also shown in Fig. 14. The rounding of the shock front arises from the volumetric viscoelasticity in the GAP model. Clearly, the GAP model compares well to the experiments. The exception is the release wave portion (the velocity drop after about 5 µs) of the velocity profiles. Barker and Hollenbach have conjectured that while no Hugoniot Elastic Limit (HEL) is observed at the shock front (an HEL is indicative of the onset of plastic flow) perhaps the shoulder observed in release portion of the wave might be explained by the onset of plastic flow. Parenthetically, these authors gave both supporting and opposing arguments for plastic flow being the cause of the shoulder. The GAP model contains plastic flow, *via* the HFS model, but does not capture the shoulder (Fig. 14), thus the present GAP HFS model does not support the plastic flow conjecture of Barker and Hollenbach (1970). Another explanation of the shoulder should be considered. Rabinowitz and coworkers (Rabinowitz, *et al.,* 1970) observed that a crossover in behavior occurs in PMMA between 3 and 4 kbar. The crossover is apparent in the maximum shear stress supported by PMMA, and is a change from yield to fracture behavior. Because this pressure is very near the stress where the shoulder first appears in Fig. 14, this may provide an alternative explanation for the shoulder. Future work, where damage is included in the GAP model, might help to resolve this issue.

Figure 15 shows the GAP model velocity profile predictions as the shock propagates deeper and deeper into the PMMA target. "Gauge" positions where the particle velocity is calculated are listed on the figure. Clearly, at 450 m/s, the wave continues to evolve, indicative of a non-steady wave. In the GAP model the cause of the evolving wave is the non-equilibrium feature of the model. The experiments of Barker and Hollenbach (1970) and Schuler (1970) report quantitatively similar behavior.

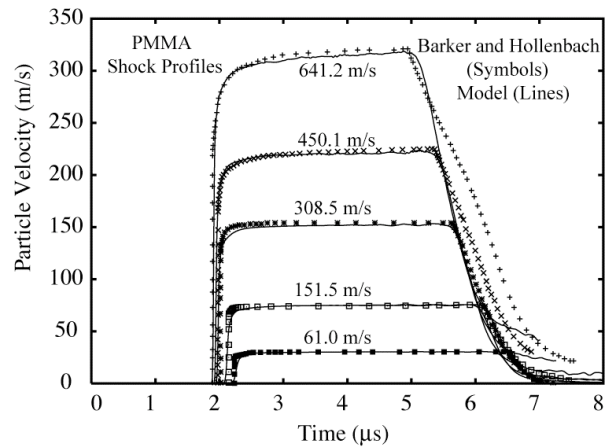

Figure 14. PMMA velocity profiles of Barker and Hollenbach (1970) (points) and from the GAP model (lines). Rounding in the profiles provide restrictions on the values used for $B_T^{(4)}$, $B_T^{(5)}$, and $B_T^{(6)}$ in Table 4. Impact velocities are listed.

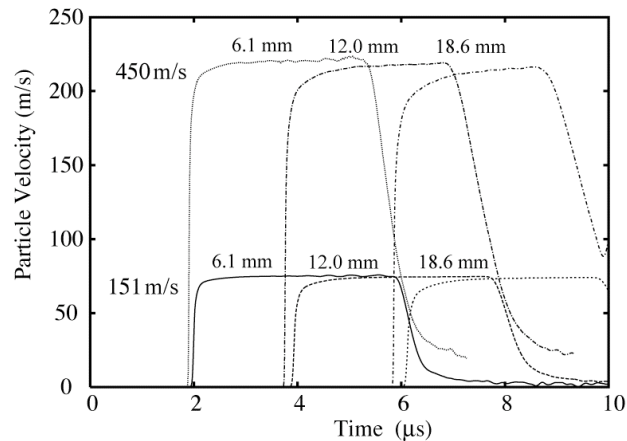

Figure 15. PMMA velocity profiles calculated from the GAP model for two different impact velocities. Positions, measured from the impact surface, are given for the calculated profiles within the PMMA target.

We close this section by taking a closer look at the evolving wave structure due to non-equilibrium physics. This is important in both the calculation of our theoretical shock Hugoniot, as well as in the experimental Hugoniot of Barker and Hollenbach (1970) and Schuler (1970). In the experiments it was noted that the peak stress continues to evolve as the shock propagates deeper and deeper into the PMMA target, until an apparent steady state is reached. At that position, the equilibrium state of the shocked PMMA was declared, and the corresponding Hugoniot point was then determined. Figure 16 shows the corresponding GAP calculation for the 308.5 m/s impact. Clearly the stress continues to relax as the shock propagates deeper into the target, in accord with the experiment, and that a steady state seems to be reached. In Fig. 16, the different curves correspond to approximately spaced locations within the PMMA target beginning at 0.4 mm and ending at 13 mm. A stress of 0.54 GPa is then recorded as the Hugoniot stress. With the corresponding particle velocity (at that location), the Hugoniot for that impact speed is determined. The locus of such points are shown in Fig. 17, along with the experimental Hugoniot. Clearly the agreement with experiment is now very good, once again demonstrating that the inclusion of the non-equilibrium EOS to the equilbrium EOS is what is required to correct the poor agreement shown in Fig. 5.

It is clear that much more volumetric relaxation is possible in PMMA (independent of our theoretical conjecture, recall Sane and Knauss's (2001) dynamic bulk response shown in Fig. 13). Our calculation shows that by the time the release waves have arrived in Fig. 16, the volumetric strain rate has relaxed from about $10^7$ $s^{-1}$ at shock arrival, to about $10^4$ $s^{-1}$ at 13 mm. Pragmatically, excluding the fact that the high pressure itself will slow the relaxations, the time that will be required to reach true equilibrium is of coarse huge and will have little practical consequence. The important point is that by including the non-equilibrium features, one now can reliably obtain the correct shock arrival times and pressure.

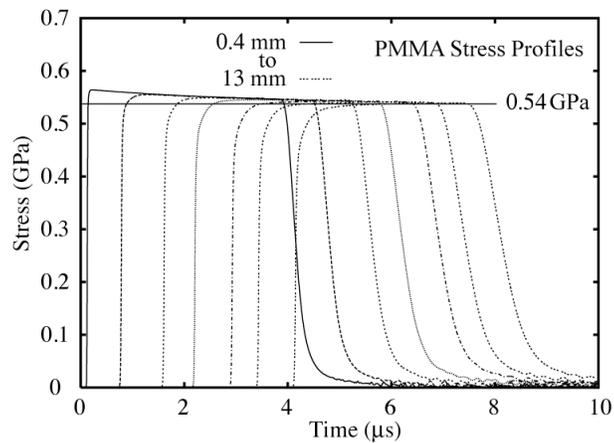

Figure 16. PMMA stress profiles calculated from the GAP model for an impact velocity of 308.5 m/s. Locations of the calculated profiles in the PMMA target are approximately evenly spaced between 0.4 and 13 mm.

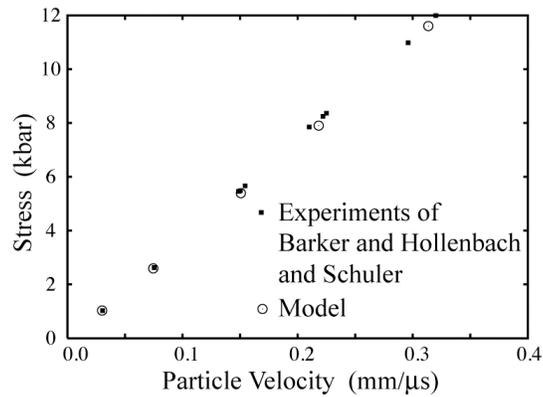

Figure 17. Hugoniot for PMMA measured by Barker and Hollenbach (1970) and Schuler (1970) and calculated from the GAP model.

**4. Conclusions**

    A continuum theory based on an *ansatz* for a non-equilibrium free energy, where both the deviatoric and the volumetric contributions to the free energy are functions of the rate of deformation, has been developed. The resulting model is called the Glassy Amorphous Polymer (GAP) model, and has been shown to accurately capture the thermo-mechanical polymer behavior ranging from equilibrium properties all the way to that observed in high-rate shock experiments. PMMA was used as a representative glassy polymer, but the model has been shown to work equally well on a host of other polymers (Clements, Unpublished results). PMMA is special in that it is one of the few polymers where sufficient experiments have been done to determine many of the parameters required by the GAP model. Of special relevance are the dynamic bulk compliance measurements of Sane and Knauss (2001). The GAP model uses a novel flow stress model called the Hierarchical Flow Stress (HFS) model. It has been shown to accurately model the stress-strain behavior of PMMA over a broad range of strain rates and temperatures, as well as capture the details of stress plateau, softening, and hardening behavior observed in glassy amorphous polymers. Because data is insufficient for the full determination of the model's non-equilibrium EOS, several plausible

conditions, called the quasi-equilibrium hypotheses, were put forth to circumvent this problem. One of the conditions, related to the non-equilibrium volume expansion coefficient, is argued to be consistent with a non-equilibrium counterpart to a Maxwell relation. The other condition is that all volumetric elements, in a generalized Maxwell representation of the volumetric response, are at a common temperature.

It is desired that this work is viewed as an initial step in a potentially successful approach to include non-equilibrium EOS properties into a thermo-mechanical framework. Clearly with measurement on non-equilibrium quantities like dynamic $c_p$, $\gamma$ or $\beta$, over a wide range of rates and temperatures, for example, some of the approximations introduced in this work might be removed. The beneficial outcome will be a more consistent theory.

As a next step in the development of this model, polymer damage is to be included. This is a critical step if the GAP model is to be useful in many impact situations. Both deviatoric and volumetric damage must be envisioned, the latter implies a coupling to both the equilibrium and non-equilibrium equations of state. A potential result of this work will shed light on the plate impact study of Barker and Hollenbach (1970) and Schuler (1970), in resolving the anomalous shoulder observed in the release portion of their measured velocity profiles in the symmetric impact of PMMA.

**Appendix A.  Fundamental Thermo-mechanics**

The goal of these appendices is to highlight the proposed non-equilibrium analysis, the underlying physical assumptions, and the polymer behavior to be captured by a glassy amorphous polymer constitutive model. To reduce complications associated with large deformation formulations (Malvern, 1969), only a small deformation theory will be described. We establish the convention that the equation of state pressure $p$ and volumetric strain $\varepsilon_v$ are positive in compression

$$\varepsilon_v > 0, \ p > 0 \quad \text{in compression.} \tag{A.1}$$

In the case of small deformations, the rate of deformation tensor $D_{ij}$ is approximately equal to the time derivative of the strain tensor $\varepsilon_{ij}$, from which the tensor trace yields

$$D_{kk} = \frac{\partial v_k}{\partial x_k} \approx \frac{d}{dt}\left(\frac{\partial u_k}{\partial X_k}\right) = \dot{\varepsilon}_{kk}. \tag{A.2}$$

Here the standard notation for the components of the velocity $v_k$, displacement $u_k$, material coordinate $X_k$, and spatial coordinate $x_k$ are used (Malvern, 1969). Letting $\sigma_{ij}$ denote the elements of the Cauchy stress tensor, for small deformations the stress power input applied to the system $\sigma_{ij} D_{ij}$ can thus be approximated as $\sigma_{ij}\dot{\varepsilon}_{ij}$. Under the approximation of small deformation mass, momentum, and energy conservation are expressed by

$$\dot{\rho} - \rho \dot{\varepsilon}_v = 0, \tag{A.3}$$

$$\frac{\partial \sigma_{ji}}{\partial x_j} = \rho \dot{v}_i, \tag{A.4}$$

$$\rho \dot{u} = \sigma_{ij}\dot{\varepsilon}_{ij} + \rho r - \frac{\partial q_k}{\partial x_k}, \tag{A.5}$$

where, the $\rho = 1/\upsilon$ is the mass density, $\upsilon$ is the specific volume, $u$ is the specific internal energy, and the last two terms in Eq. (A.5) represent the heat production per unit mass, $r$, and the heat flux leaving the system $\partial q_k / \partial x_k$, per unit mass. There are potentially two sources of heat flux, the heat transported to he surrounding material (the external heat flux) and an internal heat flux exchanged within the internal elements of the model (see Appendix E). This appendix discusses only with the external heat flux.

Substituting the deviatoric stress $s_{ij}$ and strain rate $\dot{e}_{ij}$, defined by

$$s_{ij} \equiv \sigma_{ij} - p_m \delta_{ij} \equiv \sigma_{ij} - \frac{1}{3}\sigma_{kk}\delta_{ij} = \sigma_{ij} + p\delta_{ij}, \tag{A.6}$$

and

$$\dot{e}_{ij} \equiv \dot{\varepsilon}_{ij} - \frac{1}{3}\dot{\varepsilon}_{kk}\delta_{ij} = \dot{\varepsilon}_{ij} + \frac{1}{3}\dot{\varepsilon}_{\upsilon}\delta_{ij}, \tag{A.7}$$

Eq. (A.5) becomes

$$\rho \dot{u} = s_{ij}\dot{e}_{ij} + p\dot{\varepsilon}_{\upsilon} + \rho r - \frac{\partial q_k}{\partial x_k}. \tag{A.8}$$

or

$$\rho \dot{u} = s_{ij}\dot{e}_{ij} - \rho p\dot{\upsilon} + \rho r - \frac{\partial q_k}{\partial x_k} \tag{A.9}$$

Note that $\varepsilon_{ij}$ in Eq. (A.7) is the total strain, equal to elastic plus inelastic strain contributions.

The Clausius-Duhem form (Christensen, 1971; Malvern, 1969) of the second law is used to introduce the internal entropy production rate $\gamma_D$:

$$\rho T \gamma_D = \rho T \dot{s} - \rho r + \frac{\partial q_k}{\partial x_k} - \frac{1}{T}q_k \frac{\partial T}{\partial x_k} \geq 0, \tag{A.10}$$

where $s$ is the specific entropy. In the present work fast processes are considered, which occur over short time durations for which heat conduction to and from external points can be ignored. Moreover heat source terms are omitted in the present analysis. In the event that such terms are needed, it is clear how to formally include them. Within these constraints the internal entropy production will come only from irreversible inelastic contributions (plasticity, viscoelasticity, damage, as examples), $\rho T \gamma_D = \rho T \dot{s}$. Finally, replacing the specific internal energy with the specific Helmholtz free energy, $a = u - Ts$, and using energy conservation Eq. (A.9) yields the final desired expression:

$$\rho T \gamma_D = \rho T \dot{s} = -\rho \dot{a} - \rho s \dot{T} + s_{ij} \dot{e}_{ij} - \rho p \dot{v} \geq 0. \qquad (A.11)$$

**Appendix B. Description of the GAP model**

Any useful computational model must be numerically tractable plus require input information (*i.e.*, model parameterization data) readily measurable. For polymers in non-equilibrium situations, the latter condition puts an important constraint and limitation on the model developer. In this appendix we put forth a formal theory. However, due to a lack of experimental information, the resulting model requires simplification. For glassy polymers, we argue that reasonable approximations (Appendix D) can be made in the theory that somewhat mitigates this problem. We begin by summarizing the theory.

Our ansatz for the specific Helmholtz free energy is a sum of equilibrium and non-equilibrium terms

$$\rho a(\varepsilon_v^E, e_{ij}^E, T; \{\dot{\varepsilon}_v^E, \dot{e}_{ij}^e, \dot{T}\}) = \rho a^{(0)}(v, e_{ij}^E, T) + \frac{1}{2}\left(\sum_{n=1}^{N} \frac{s_{ij}^{(n)} s_{ij}^{(n)}}{2\mu^{(n)}} + \sum_{m=1}^{M} \frac{p^{(m)} p^{(m)}}{B_T^{(m)}}\right), \qquad (B.1)$$

where $a^{(0)}(\upsilon, e_{ij}^E, T)$ is the equilibrium specific Helmholtz free energy and the second term is our approximation for the non-equilibrium contributions for both deviatoric and volumetric parts of the free energy. We will call the $n = 1,...,N$ deviatoric and $m = 1,...,M$ volumetric terms as the deviatoric and volumetric *elements* in the model, respectively. Here $\upsilon$ is the specific volume and $e_{ij}^E$ is the elastic part of the deviatoric strain. The curly brackets indicate that the total specific Helmholtz free energy is a functional of the strain and temperature history. The proper independent variables for the Helmholtz free energy enter through a judicious choice of the non-equilibrium contributions to the pressure $p^{(m)}$ and deviatoric stress $s_{ij}^{(n)}$ as

$$p^{(m)}(t) = \int_0^t dt_1 e^{\bar{\phi}^{(m)}(t_1) - \bar{\phi}^{(m)}(t)} B_T^{(m)}(t_1) \dot{\varepsilon}_\upsilon(t_1)$$
$$+ \int_0^t dt_1 e^{\bar{\phi}^{(m)}(t_1) - \bar{\phi}^{(m)}(t)} \beta^{(m)}(t_1) B_T^{(m)}(t_1) \dot{T}(t_1) \qquad (B.2)$$

and

$$s_{ij}^{(n)}(t) = \int_0^t dt_1 e^{\phi^{(n)}(t_1) - \phi^{(n)}(t)} 2\mu^{(n)}(t_1) \dot{e}_{ij}^{\eta(n)}(t_1). \qquad (B.3)$$

In Eq. (B.2), $B_T^{(m)}(t)$ is the $m^{th}$ isothermal bulk relaxation moduli, $\beta^{(m)}(t)$, is defined formally as the $m^{th}$ element of nonequilibrium volumetric expansion coefficient (also sometimes called the volumetric expansivity), $T(t)$ is the temperature, $\mu^{(n)}(t)$ is the $n^{th}$ shear relaxation modulus, and $d\varepsilon_\upsilon = -d\upsilon/\upsilon$. $\phi^{(n)}(t) = \phi^{(n)}\left(\tau^{(n)}(t)\right)$ and $\bar{\phi}^{(m)}(t) = \bar{\phi}^{(m)}\left(\bar{\tau}^{(m)}(t)\right)$, defined in Appendix C, are functions of the temperature-dependent shear and volumetric relaxation times $\tau^{(n)}(t)$ and $\bar{\tau}^{(m)}(t)$, respectively. The strain rate $\dot{e}_{ij}^{\eta(n)} \equiv \dot{e}_{ij} - \dot{e}_{ij}^{P(n)}$ is described in detail in Appendix C, but differs from the total deviatoric strain rate by the inclusion of deviatoric plasticity. The moduli in Eq. (B.2) and (B.3) will be taken to depend on volumetric compression. The temperature-dependence of moduli comes through the $\tau^{(n)}(t)$ and $\bar{\tau}^{(m)}(t)$ as described in Appendix C.

There is already an important assumption inherent in Eq. (B.1) and (B.2) regarding the temperature. While a temperature $T(t)$ can be defined for the free energy, the assumption that each of the $m = 0,..., M$ elements are at the same temperature is a hypothesis that we call the first condition of the *quasi-equilibrium hypothesis*. That this is a plausible hypothesis is derived from notion that heat may be transported rapidly between the elements, *i.e.*, the various modes of excitation, such that a common temperature is maintained at all times. This clearly need not be the case, however, and in a more sophisticated theory $T(t)$ might represent the average temperature of the individual element temperatures. Such a step might be expected to be important in extending the present theory to include non-equilibrium situations that arise from purely thermal agitations (Wunderlich, 2007; Garden, 2007).

Another important point occurs because thermodynamic quantities (*i.e.*, quantities that are derivable from derivatives of the free energy) are already explicit in Eq. (B.2), these quantities must satisfy certain thermodynamic constraining equations. This will be discussed below when a non-equilibrium Maxwell relation is investigated.

The differential of Eq. (B.2) and (B.3) produces equations for each

$$dp^{(m)} = B_T^{(m)} d\varepsilon_v + \beta^{(m)} B_T^{(m)} dT - \frac{p^{(m)}}{\bar{\tau}^{(m)}} dt ,\qquad (B.4)$$

and

$$ds_{ij}^{(n)} = 2\mu^{(n)} de_{ij}^{\eta(n)} - \frac{s_{ij}^{(n)}}{\tau^{(n)}} dt = 2\mu^{(n)} \left( de_{ij} - de_{ij}^{P(n)} \right) - \frac{s_{ij}^{(n)}}{\tau^{(n)}} dt .\qquad (B.5)$$

Similar expressions to Eq. (B.4) and (B.5) exist, but with the omission of the relaxation terms,

$$dp^{(0)} = B_T^{(0)} d\varepsilon_v + \beta^{(0)} B_T^{(0)} dT , \tag{B.6}$$

$$ds_{ij}^{(0)} = 2\mu^{(0)} de_{ij}^{E(0)} = 2\mu^{(0)} \left( de_{ij} - de_{ij}^{P(0)} \right), \tag{B.7}$$

for the equilibrium terms. Use of Eq. (C.5) and (C.9) described in Appendix C are needed to derive Eq. (B.4) and (B.5). Note that a caveat must be issued at this point. Recall that the index 0 denotes the equilibrium values. For example, $B_T^{(0)}$ is to be determined by taking the volumetric derivatives of $a^{(0)}(v, e_{ij}^E, T)$ (see Eq. 10, for example). As such, it has a temperature and volume dependence derived from the equilibrium free energy, and is different from the simple parameterization assigned to $B_T^{(m)}$ for $m \geq 1$, for which we take a simple polynomial approximation

$$B_T^{(m)}(t) = B_T^{(m)} \left( 1 + \chi_1 \varepsilon_v(t) + \chi_2 \varepsilon_v^2(t) \right) \quad m \geq 1 , \tag{B.8}$$

where $\chi_1$ and $\chi_2$ are zero in expansion. To account for the volumetric dependence in the shear response a similar parameterization is used:

$$\mu^{(n)}(t) = \mu_o^{(n)} \left( 1 + \chi_3 \varepsilon_v(t) \right) \quad n \geq 0 . \tag{B.9}$$

Regarding Eq. (B.9), we note that numerous experiments have shown that polymers exhibit volumetric dependent (*i.e.*, pressure dependent) shear moduli (Sauer, 1977; Sauer *et al*.,

1970; Pae and Bhateja, 1975). In Eq. (B.8) and (B.9), $\chi_1$, $\chi_2$ and $\chi_3$ are treated as constant material parameters.

The differential of the Eq. (B.1) yields

$$\rho da = \rho da^{(0)} + \sum_{m=1}^{M} \frac{p^{(m)}}{B_T^{(m)}} dp^{(m)} + \sum_{n=1}^{N} \frac{s_{ij}^{(n)}}{2\mu^{(n)}} ds_{ij}^{(n)}$$
$$- \frac{1}{2} \sum_{m=1}^{M} \left( \frac{p^{(m)}}{B_T^{(m)}} \right)^2 \left( \frac{\partial B_T^{(m)}}{\partial v} \right)_T dv \qquad (B.10)$$
$$- \frac{1}{2} \sum_{n=1}^{N} \left( \frac{s_{ij}^{(n)}}{2\mu^{(n)}} \right)^2 \left( \frac{\partial (2\mu^{(n)})}{\partial v} \right)_T dv$$

It is easily shown that in the small strain approximation, the derivative of the $\rho$ multiplying the free energy in Eq. (B.1) must not be included. This is readily shown for simple free energies, but holds true also for the current model. Given the free energy *ansatz*, our first approximation of this appendix is to neglect the last two terms in Eq. (B.10), being of second-order in the volumetric and deviatoric strains, while the second and third terms are of first order. Inserting Eq. (B.4) and (B.5) into Eq. (B.10) results in

$$\rho da = \rho da^{(0)} - \sum_{m=1}^{M} \frac{p^{(m)}}{v} dv + \sum_{m=1}^{M} \beta^{(m)} p^{(m)} dT + \sum_{n=1}^{N} s_{ij}^{(n)} de_{ij}^{\eta(n)}$$
$$- \sum_{m=1}^{M} \frac{p^{(m)} p^{(m)}}{B_T^{(m)} \overline{\tau}^{(m)}} dt - \sum_{n=1}^{N} \frac{s_{ij}^{(n)} s_{ij}^{(n)}}{2\mu^{(n)} \tau^{(n)}} dt \qquad (B.11)$$

The equilibrium contributions to the pressure, deviatoric stress, and specific entropy are

$$p^{(0)} = -\frac{\partial a^{(0)}}{\partial v}\bigg)_{e_{ij}^{E(0)},T}$$

$$s_{ij}^{(0)} = \rho \frac{\partial a^{(0)}}{\partial e_{ij}^{E(0)}}\bigg)_{\varepsilon_v,T} \quad (B.12)$$

$$s^{(0)} = -\frac{\partial a^{(0)}}{\partial T}\bigg)_{v,e_{ij}^{E(0)}}$$

Recalling Eq. (A.11), *i.e.*,

$$\rho T ds = -\rho da - \rho s dT + s_{ij} de_{ij} - \rho p dv \geq 0, \quad (B.13)$$

substituting Eq. (B.11) into Eq. (B.13), and equating the prefactors of $dv$, $de_{ij}$, and $dT$ to zero (by the usual arguments of Collman and Noll (1963), results in the pressure equation

$$p = p^{(0)} + \sum_{m=1}^{M} p^{(m)}, \quad (B.14)$$

the stress deviator equation

$$s_{ij} = s_{ij}^{(0)} + \sum_{n=1}^{N} s_{ij}^{(n)}, \quad (B.15)$$

and the specific entropy

$$s = s^{(0)} + \sum_{m=1}^{M} s^{(m)} = s^{(0)} - v \sum_{m=1}^{M} \beta^{(m)} p^{(m)}. \quad (B.16)$$

Finally, the remaining terms can be equated to the internal dissipation,

$$\rho T \gamma_D = \rho T \frac{ds}{dt} = \sum_{m=1}^{M} \frac{p^{(m)} p^{(m)}}{B_T^{(m)} \bar{\tau}^{(m)}} + \sum_{n=1}^{N} \frac{s_{ij}^{(n)} s_{ij}^{(n)}}{2\mu^{(n)} \tau^{(m)}} + \sum_{n=0}^{N} s_{ij}^{(n)} \dot{e}_{ij}^{P(n)} \geq 0 \,. \tag{B.17}$$

The first two terms are the volumetric and deviatoric viscoelastic dissipations while the final term arises from devatoric plasticity.

The next step in the analysis is to examine higher-order thermodynamic relations, beginning with the well-known equilibrium relations. In equilibrium, all the standard thermodynamic definitions and relations apply:

$$T \frac{\partial s^{(0)}}{\partial T} \bigg)_v = c_v^{(0)} \qquad T \frac{\partial s^{(0)}}{\partial T} \bigg)_p = c_p^{(0)} \qquad v \frac{\partial s^{(0)}}{\partial v} \bigg)_T = \gamma^{(0)} c_v^{(0)} \tag{B.18-20}$$

$$v \frac{\partial T}{\partial v} \bigg)_s = -\gamma^{(0)} T \qquad v \frac{\partial p^{(0)}}{\partial v} \bigg)_T = -B_T^{(0)} \qquad \frac{\partial p^{(0)}}{\partial T} \bigg)_v = \beta^{(0)} B_T^{(0)} \tag{B.21-23}$$

and

$$\frac{1}{v} \frac{\partial v}{\partial T} \bigg)_p = \beta^{(0)} \,. \tag{B.24}$$

From these relations and the Maxwell relation

$$\frac{\partial s^{(0)}}{\partial v} \bigg)_T = \frac{\partial p^{(0)}}{\partial T} \bigg)_v \tag{B.25}$$

it follows that

$$\gamma^{(0)} c_v^{(0)} = v \beta^{(0)} B_T^{(0)} \tag{B.26}$$

and is easily shown that

$$c_p^{(0)} - c_v^{(0)} = T\beta^{(0)}\gamma^{(0)}c_v^{(0)}.\tag{B.27}$$

The non-equilibrium counterparts of these equations are now investigated. We first focus on rigorous relations and then refine the discussion to include approximations that lead to a tractable theory, given that experimental data is limited. The analysis will begin with the entropy equation, Eq. (B.16). Recalling that $d\varepsilon_v = -dv/v$, and identifying the viscoelastic differential volumetric strain from Eq. (B.17) as $d\varepsilon_v^{VE(m)} = p^{(m)}/\left(B_T^{(m)}\overline{\tau}^{(m)}\right)dt$, then from Eq. (B.4)

$$dp^{(m)} = B_T^{(m)}d\varepsilon_v^{E(m)} + \beta^{(m)}B_T^{(m)}dT,\tag{B.28}$$

where

$$d\varepsilon_v^{E(m)} = d\varepsilon_v - d\varepsilon_v^{VE(m)},\tag{B.29}$$

one deduces

$$\left.\frac{\partial p^{(m)}}{\partial \varepsilon_v^{E(m)}}\right)_{T,e_{ij}^{E(n)}} = B_T^{(m)},\tag{B.30}$$

and

$$\left.\frac{\partial p^{(m)}}{\partial T}\right)_{\varepsilon_v^{E(m)},e_{ij}^{E(n)}} = \beta^{(m)}B_T^{(m)}.\tag{B.31}$$

We finally state our formal expression for $\beta^{(m)}$

$$\beta^{(m)} = -\left.\frac{\partial \varepsilon_v^{E(m)}}{\partial T}\right)_{p^{(m)},e_{ij}^{E(n)}}.\tag{B.32}$$

It is clear that Eqs. (B.30)-(B.32) are non-equilibrium counterparts of the equilibrium expressions given by Eqs. (B.22)-(B.24).

To construct a Maxwell relaxation, Eq. (B.11) can be first be written as

$$\rho da = \rho da^{(0)} + \sum_{m=1}^{M} p^{(m)} d\varepsilon_v^{E(m)} + \sum_{n=1}^{N} s_{ij}^{(n)} de_{ij}^{E(n)} - \rho \sum_{m=1}^{M} s^{(m)} dT \qquad (B.33)$$

from which the Maxwell relation follows:

$$-\left.\frac{\partial s^{(m)}}{\partial \varepsilon_v^{E(m)}}\right)_{T,e_{ij}^{E(n)}} = v \left.\frac{\partial p^{(m)}}{\partial T}\right)_{\varepsilon_v^{E(m)},e_{ij}^{E(n)}}. \qquad (B.34)$$

We will return to this expression, but first note that

$$\gamma^{(m)} c_v^{(m)} = -\left.\frac{\partial s^{(m)}}{\partial \varepsilon_v^{E(m)}}\right)_{T,e_{ij}^{E(n)}} \qquad (B.35)$$

results in

$$\gamma^{(m)} c_v^{(m)} = v \beta^{(m)} B_T^{(m)}, \qquad (B.36)$$

which is a non-equilibrium counterpart to Eq. (B.26).

Next multiplying

$$c_v^{(m)} = T \left.\frac{\partial s^{(m)}}{\partial T}\right)_{\varepsilon_v^{E(m)},e_{ij}^{E(n)}} \qquad (B.37)$$

by $dT$, and Eq. (B.35) by $Td\varepsilon_v^{E(m)}$, and adding the results together yield

$$-T\gamma^{(m)} c_v^{(m)} d\varepsilon_v^{E(m)} + c_v^{(m)} dT = Tds^{(m)} \qquad (B.38)$$

where it was recognized that

$$\left.\frac{\partial s^{(m)}}{\partial \varepsilon_v^{E(m)}}\right)_{T,e_{ij}^{E(n)}} d\varepsilon_v^{E(m)} + \left.\frac{\partial s^{(m)}}{\partial T}\right)_{\varepsilon_v^{E(m)},e_{ij}^{E(n)}} dT = ds^{(m)} \qquad (B.39)$$

The first condition of the *quasi-equilibrium hypothesis* demands a common temperature among the elements. To achieve this a reversible heat exchange $dQ_{(q)}^{(m)}$ term, coming from heat added to element $m$ from the remaining elements is to be added to the entropy increment (the author is indebt to J. N. Johnson for discussions on this point),

$$-\rho T \gamma^{(m)} c_v^{(m)} d\varepsilon_v^{E(m)} + c_v^{(m)} dT = T ds^{(m)} + \sum_{\substack{q=0 \\ q \neq m}}^{M} dQ_q^{(m)} . \qquad (B.40)$$

However, upon summing over all $m = 0, ..., M$ elements and noting that $dQ_{(q)}^{(m)} = -dQ_{(m)}^{(q)}$, we find the last term cancels and we arrive at an expression for the temperature increment

$$\begin{aligned}
dT &= \left[\sum_{m=0}^{M} c_v^{(m)}\right]^{-1} \left\{ \sum_{m=0}^{M} T ds^{(m)} + T \sum_{m=0}^{M} \gamma^{(m)} c_v^{(m)} d\varepsilon_v^{E(m)} \right\} \\
&= \frac{1}{c_v^{eff}} \left\{ \sum_{m=0}^{M} T ds^{(m)} + T \sum_{m=0}^{M} \gamma^{(m)} c_v^{(m)} d\varepsilon_v - T\upsilon \sum_{m=0}^{M} \beta^{(m)} \frac{p^{(m)}}{\overline{\tau}^{(m)}} dt \right\}.
\end{aligned} \qquad (B.41)$$

or

$$dT = \frac{1}{c_v^{eff}} \sum_{m=0}^{M} T ds^{(m)} - \rho T \gamma_{eff} d\upsilon - \frac{T\upsilon}{c_v^{eff}} \sum_{m=0}^{M} \beta^{(m)} \frac{p^{(m)}}{\overline{\tau}^{(m)}} dt , \qquad (B.42)$$

where an effective Grüneisen gamma coefficient $\gamma_{eff}$ and specific heat have been introduced:

$$\gamma_{\text{eff}} = \frac{1}{c_v^{\text{eff}}} \sum_{m=0}^{M} \gamma^{(m)} c_v^{(m)}, \tag{B.43}$$

where

$$c_v^{\text{eff}} = \sum_{m=0}^{M} c_v^{(m)}. \tag{B.44}$$

Next, using Eq. (B.1) and (B.2) together implies that the thermodynamic quantities explicit in Eq. (B.2) must be constrained. To see this, return to the Maxwell relation Eq. (B.34) and substitute in $s^{(m)} = -\beta^{(m)} \upsilon p^{(m)}$. Using Eq. (B.30) and (B.31) immediately results in

$$\left. \frac{\partial \beta^{(m)} \upsilon}{\partial \varepsilon_v^{E(m)}} \right)_{T, e_{ij}^{E(m)}} = 0, \tag{B.45}$$

as a necessary condition to be satisfied to preserve the Maxwell relation. A further discussion of Eq. (B.45) is provided in Appendix D. In the remainder of this appendix we examine its consequences.

First, from the identity

$$\left. \frac{\partial f}{\partial T} \right)_p - \left. \frac{\partial f}{\partial T} \right)_\upsilon = \left. \frac{\partial f}{\partial \upsilon} \right)_T \left. \frac{\partial \upsilon}{\partial T} \right)_p, \tag{B.46}$$

and choosing $f \to \beta^{(m)} \upsilon$, $p \to p^{(m)}$, and $\upsilon \to \varepsilon_v^{E(m)}$ we deduce

$$\left.\frac{\partial \beta^{(m)} \upsilon}{\partial T}\right)_{p^{(m)}} - \left.\frac{\partial \beta^{(m)} \upsilon}{\partial T}\right)_{\varepsilon_\upsilon^{E(m)}} = \left.\frac{\partial \beta^{(m)} \upsilon}{\partial \varepsilon_\upsilon^{E(m)}}\right)_T \left.\frac{\partial \varepsilon_\upsilon^{E(m)}}{\partial T}\right)_{p^{(m)}} = 0 \qquad (B.47)$$

Taking the temperature derivative of the entropy, at constant elastic volume, we identify the $m = 1,...,M$ elements of the non-equilibrium specific heat as

$$c_\upsilon^{(m)} = T \left.\frac{\partial s^{(m)}}{\partial T}\right)_{\varepsilon_\upsilon^{E(m)}, e_{ij}^{E(n)}}$$

$$= -T \left.\frac{\partial \beta^{(m)} \upsilon}{\partial T}\right)_{\varepsilon_\upsilon^{E(m)}} p^{(m)} - T\upsilon \beta^{(m)} \left.\frac{\partial p^{(m)}}{\partial T}\right)_{\varepsilon_\upsilon^{E(m)}}$$

$$= -T \left.\frac{\partial \beta^{(m)} \upsilon}{\partial T}\right)_{\varepsilon_\upsilon^{E(m)}} p^{(m)} - T\upsilon \left(\beta^{(m)}\right)^2 B_T^{(m)} . \qquad (B.48)$$

Likewise

$$c_p^{(m)} = -T \left.\frac{\partial \beta^{(m)} \upsilon}{\partial T}\right)_{p^{(m)}, e_{ij}^{E(n)}} p^{(m)} \qquad (B.49)$$

Equations (B.47), (B.48) and (B.49) can be combined to yield

$$c_p^{(m)} - c_\upsilon^{(m)} = T\upsilon \left(\beta^{(m)}\right)^2 B_T^{(m)} , \qquad (B.50)$$

and together with Eq. (B.26) is the non-equilibrium counterpart to Eq. (B.27).

Aside from discussing non-isothermal viscoelasticity (Appendix C), this completes the formal analysis. Approximations must be introduced that are compatible with available experimental information. This is the objective of Appendix D.

**Appendix C. Nonisothermal Viscoelasticity**

The inelastic deformation and adiabatic heating at high rates of deformation generate a local temperature rise in a polymer according to Eq. (B.42). Because both the shear and bulk moduli are strongly temperature dependent in polymers, it is important to account for the local temperature change. This implies that a non-isothermal solution of the viscoelastic equations must be used. Beginning with basic differential equation for deviatoric viscoelasticity

$$\dot{s}_{ij}^{(n)}(t) = 2\mu^{(n)}(t)\dot{e}_{ij}^{\eta(n)}(t) - s_{ij}^{(n)}(t)/\tau^{(n)}(t), \quad n \geq 1 \quad (C.1)$$

where $\dot{e}_{ij}^{\eta(n)}(t)$ is distinguished from the total deviatoric strain rate $\dot{e}_{ij}(t)$ by the presence of inelastic strains coming from plasticity (and damage in future calculations), for example

$$\dot{e}_{ij}^{\eta(n)}(t) = \dot{e}_{ij}(t) - \dot{e}_{ij}^{P(n)}(t). \quad (C.2)$$

By equating

$$\dot{e}_{ij}^{VE(n)} = \frac{s_{ij}^{(n)}}{2\mu^{(n)}\tau^{(n)}} \quad (C.3)$$

with the deviatoric viscoelastic strain rate, an elastic strain rate can also be defined

$$\dot{e}_{ij}^{E(n)}(t) = \dot{e}_{ij}^{\eta(n)}(t) - \dot{e}_{ij}^{VE(n)} = \dot{e}_{ij}(t) - \dot{e}_{ij}^{P(n)}(t) - \dot{e}_{ij}^{VE(n)} \quad (C.4)$$

By introducing

$$\dot{\phi}^{(n)}(t) \equiv 1/\tau^{(n)}(t), \quad (C.5)$$

the integral of Eq. (C.1) is immediately determined, which is Eq. (B.3),

$$s_{ij}^{(n)}(t) = \int_0^t dt_1 e^{\phi^{(n)}(t_1) - \phi^{(n)}(t)} 2\mu^{(n)}(t_1) \dot{e}_{ij}^{\eta(n)}(t_1), \quad (C.6)$$

where

$$\phi^{(n)}(t) = \int_0^t \frac{dt'}{\tau^{(n)}(t')}. \quad (C.7)$$

In Eq. (C.1), the time dependence of $\mu^{(n)}(t)$ is a result of the implicit time dependence of the volumetric strain through Eq. (B.9) while the time dependence of the relaxation time comes from the temperature history: $\tau^{(n)}(t) = \tau^{(n)}(T(t))$.

The analog of this analysis holds for the volumetric viscoelastic response, Eq. (B.2) and (B.4), but with the Eq. (C.7) replaced with

$$\bar{\phi}^{(m)}(t) = \int_0^t \frac{dt'}{\bar{\tau}^{(m)}(t')}, \quad (C.8)$$

and Eq. (C.5) replaced with

$$\dot{\bar{\phi}}^{(m)}(t) \equiv 1/\bar{\tau}^{(m)}(t). \quad (C.9)$$

Also in the present work, only devatoric plasticity is considered ($\dot{\varepsilon}_v^P = 0$) and thus

$$\dot{\varepsilon}_v^{E(m)} = \dot{\varepsilon}_v - \dot{\varepsilon}_v^{VE(m)}. \tag{C.10}$$

**Appendix D. Working Assumptions and Approximations**

This appendix begins with a discussion of $\beta^{(m)}$, *i.e.*,

$$\beta^{(m)} = -\frac{\partial \varepsilon_v^{E(m)}}{\partial T}\bigg)_{p^{(m)}, e_{ij}^{E(n)}}. \tag{D.1}$$

Holding the $m^{th}$ element of the pressure, $p^{(m)}$, fixed in a dilatometric volume-temperature measurement is not experimentally feasible. Nevertheless, the volumetric expansion coefficient of an amorphous polymer is expected to have non-equilibrium (rate) dependence, and thus Eq. (D.1) presents a problem. Ferry (1980) discusses the rate dependent behavior of $\upsilon$ at some length and refers to the extensive work of Kovacs (1958) who showed that cooling a polymer at different rates, from starting temperatures in the vicinity of the glass transition, lead to a family of specific volume curves, as depicted in Fig. 4. Although the pressure is not held fixed in Kovacs' measurements, *i.e.*, the slope is not identically $\beta\upsilon$, below the glass transition the equilibrium volumetric expansion coefficient differs from the non-equilibrium one observed at non-zero rates. Moreover, while Ferry's discussion is relevant to the regime of volumetric creep, the concepts can intuitively be extended to high rates. Thus we claim that $\beta^{(m)}$ should not be approximated as zero in our expressions.

Because $\beta^{(m)}$ appears in the present theory as always multiplied by a non-equilibrium quantity, for example as a product with $p^{(m)}$ in the entropy $s^{(m)} = \upsilon \beta^{(m)} p^{(m)}$ in Eq. (B.16) and (B.42), and multiplied by $B_T^{(m)}$ in Eq. (B.2), our next approximation is to take

$$\beta^{(m)} \approx \frac{1}{\upsilon}\frac{\partial \upsilon}{\partial T}\bigg)_p \equiv \beta^{(0)}. \tag{D.2}$$

For the entropy example, Eq. (D.2) has the consequence that the important non-equilibrium relaxation processes of the entropy $s^{(m)} = \upsilon \beta^{(0)} p^{(m)}$ is only through the mechanical relaxation processes of $p^{(m)}$. Because we have omitted the non-equilibrium effects arising from $\beta^{(m)}$, we refer to Eq. (D.2) as the second condition of the *quasi-equilibrium hypothesis*.

Our next observation comes from the general behavior of the specific volume $\upsilon$. From Fig. 4 it clear that for temperatures not close to the glass transition, many polymers are quit well characterized by $\upsilon \beta^{(0)} \approx$ constant. This implies that all the derivatives of $\upsilon \beta^{(0)}$ will be small, including the one in Eq. (B.45), required to satisfy a Maxwell relation. Indeed an alternative to Eq. 6, that provides a relatively good fit to the specific volume data both below and separately above the glass transition is a linear form in temperature and pressure:

$$\upsilon(p,T) = \upsilon_{ref} + A(T - T_{ref}) + B(p - p_{ref}), \tag{D.3}$$

($A, B$ are constants) for which the second derivatives vanish. This form has been used by Weir, for example to describe polytetrafluroethylene (Weir, 1954). Our second approximation is then to take $\beta^{(0)} \upsilon$ equal to a constant, and by the given arguments, this approximation is consistent with actual glassy polymer behavior.

This has immediate consequences. From Eq. (B.36), (B.48), and (B.49)

$$c_v^{(m)} = -Tv\left(\beta^{(0)}\right)^2 B_T^{(m)}, \tag{D.4}$$

$$c_p^{(m)} = 0, \tag{D.5}$$

$$\gamma^{(m)} = -\left(T\beta^{(0)}\right)^{-1}, \tag{D.6}$$

and

$$\gamma^{(m)} c_v^{(m)} = v\beta^{(0)} B_T^{(m)}, \qquad m = 1, ..., M. \tag{D.7}$$

From these results it follows that Eq. (B.44) can be estimated as

$$\begin{aligned} c_v^{\text{eff}} &= \sum_{m=0}^{M} c_v^{(m)} = c_v^{(0)} + \sum_{m=1}^{M} c_v^{(m)} = c_v^{(0)} - Tv\beta^{(0)2}\xi(M')B_T^{(0)} \\ &= c_v^{(0)}(1 - \beta^{(0)}T\gamma^{(0)}\xi(M')) \approx c_v^{(0)} \end{aligned} \tag{D.8}$$

where $\xi(M')$, introduced in Eq. (17), is a measure of the deviation of the non-equilibrium isothermal bulk moduli relative to the equilibrium contribution,

$$\sum_{m=1}^{M'} B_T^{(m)} \equiv \xi(M')B_T^{(0)} \tag{D.9}$$

The last approximation in Eq. (D.8) follows because the $\beta^{(0)}T\gamma^{(0)}\xi$ can be estimated to be maximally on the order of a 0.1-0.3 and, being subject to rather large uncertainties due to the calculation its constituents, it is neglected. Thus, keeping only the active elements in the summation,

$$\sum_{m=0}^{M}\gamma^{(m)}c_v^{(m)} = \upsilon\beta^{(0)}\left[B_T^{(0)} + \sum_{m=1}^{M'}B_T^{(m)}\right] \tag{D.10}$$

$$= \upsilon\beta^{(0)}B_T^{(0)}\left(1+\xi(M')\right) = \gamma^{(0)}c_v^{(0)}\left(1+\xi(M')\right)$$

and finally,

$$\gamma_{eff} = \frac{1}{c_v^{eff}}\sum_{m=0}^{M}\gamma^{(m)}c_v^{(m)} = \gamma^{(0)}(1+\xi(M')) \tag{D.11}$$

This completes the discussion of the approximate solution, but one that is compatible with available experimental data.

A consequence of this work leads to an interesting expression for the entropy. From the differential of the $m^{\text{th}}$ element of the entropy:

$$ds^{(m)} = -\upsilon\beta^{(0)}dp^{(m)}$$

$$= \beta^{(0)}B_T^{(m)}d\upsilon - \upsilon\left(\beta^{(0)}\right)^2 B_T^{(m)}dT + \frac{\upsilon\beta^{(0)}p^{(m)}}{\overline{\tau}^{(m)}}dt$$

$$= \beta^{(0)}B_T^{(m)}d\upsilon - \upsilon\left(\beta^{(0)}\right)^2 B_T^{(m)}dT - \frac{s^{(m)}}{\overline{\tau}^{(m)}}dt \tag{D.12}$$

we immediately obtain a hysteric integral for the $m = 1,\ldots, M$ entropy elements

$$s^{(m)}(t) = \int_0^t dt_1 e^{\overline{\phi}^{(m)}(t_1)-\overline{\phi}^{(m)}(t)}\beta^{(0)}(t_1)B_T^{(m)}(t_1)\left\{\dot{\upsilon}(t_1) - \upsilon(t_1)\beta^{(0)}(t_1)\dot{T}(t_1)\right\}. \tag{D.13}$$

**Appendix E. Maxwell Model Mechanical Analog**

Using a system of springs and dashpots as a mechanical analog for the mechanical behavior of molecular polymer has a long history (Ferry, 1980). The benefit to such approaches is that once a spring/dashpot configuration has been decided upon, it becomes straightforward to write down the underlying equations. A problem is that these approaches leads one to lose sight of there being a real measurable relaxation spectrum - polymeric excitations are real and their decay towards equilibrium is describable by physical relaxation times that can be equated to the molecular motion - and that a Prony series representation is a simplistic discrete representation of that (nearly continuous) relaxation spectrum. With that caveat, the deviatoric and volumetric Generalized Maxwell Models (GMM) consistent with our non-equilibrium free energy are shown in Figs. E1 and E2, where each branch of the GMM corresponds to an element.

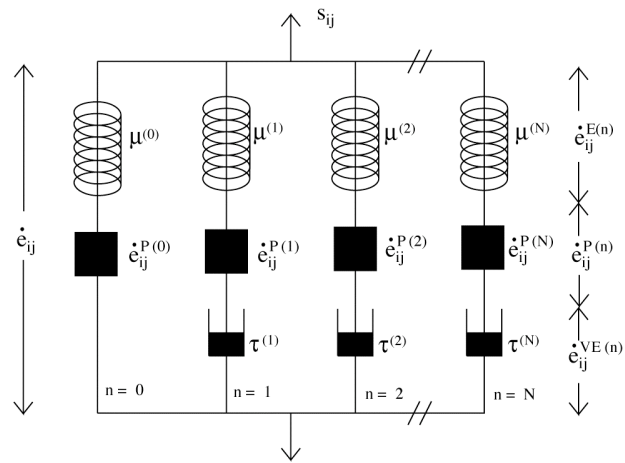

Figure E1. Deviatoric Generalized Maxwell Model representation with $n = 0,...,N$ elements.

The GMM asserts common strains (rates) in each GMM element that are equal to the applied deviatoric strain (rate). The total deviatoric stress is simply the sum of the GMM element stresses

$$\dot{e}_{ij} = \dot{s}_{ij}^{(n)} / 2\mu^{(n)} + \dot{e}_{ij}^{P(n)} + \dot{e}_{ij}^{VE(n)}, \qquad n = 1,...,N$$

$$\dot{e}_{ij} = \dot{s}_{ij}^{(0)} / 2\mu^{(0)} + \dot{e}_{ij}^{P(0)}, \qquad n = 0 \qquad (E.1)$$

$$s_{ij} = s_{ij}^{(0)} + \sum_{n=1}^{N} s_{ij}^{(n)}.$$

Using Eq. (C.3) in Eq. (E.1) results in Eq. (C.1) upon rearranging terms. The assumption of a common temperature for all GMM elements is usually implicit in GMM calculations. This point will be addressed more completely in the volumetric GMM.

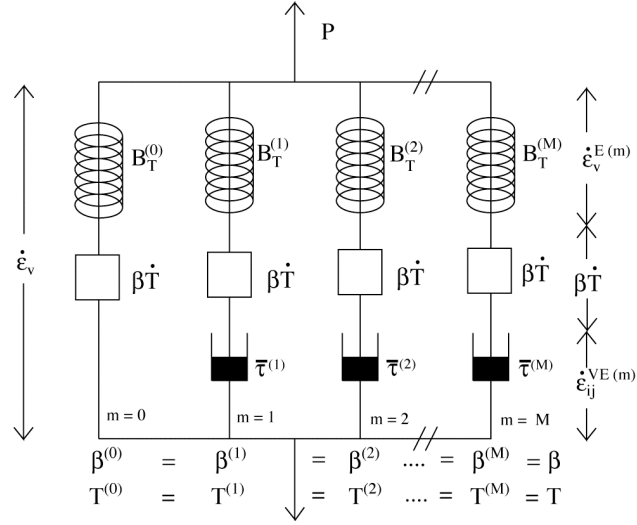

Figure E2. Volumetric Generalized Maxwell Model representation.

Similar to the deviatoric case, which results in Eq. (E.1), the volumetric equations (Eq. (B.4), (B.6) and the rate form of Eq. (B.14)) are similarly derived when our two conditions of the *quasi-equilibrium hypothesis* are invoked:

$$T^{(0)} = T^{(1)} = T^{(2)} = ... = T^{(M)} = T, \qquad (E.2)$$

and

$$\beta^{(1)} = \beta^{(2)} = ... = \beta^{(M)} = \beta^{(0)}. \qquad (E.3)$$

The details are straightforward and thus are omitted here. Rather the question of achieving a common temperature in each GMM element is pursued. Clearly, one can imagine heat being transferred rapidly between GMM elements such that a common temperature is maintained. An external heat bath (*i.e.* allowing for external heat transfer) could be included in the discussion, but for now the system will be taken as closed. Beginning then from the standard thermodynamic identity

$$c_V dT = c_V \gamma T d\varepsilon_v^E + T ds + dQ, \tag{E.4}$$

where to the irreversible entropic contribution $Tds$ (here the internal dissipation is described by Eq. (B.17)) a reversible heat flow per unit mass, $dQ$, is added, it is assumed that each GMM element must obey this identity (J. N. Johnson, Personal Communication). Then for element $m$,

$$c_V^{(m)} dT = c_V^{(m)} \gamma^{(m)} T d\varepsilon_v^{E(m)} + T ds^{(m)} + \sum_{\substack{n=1 \\ n \neq m}}^{M} dQ_n^{(m)}, \tag{E.5}$$

where $dQ_n^{(m)}$ is the heat flow (positive or negative) into element $m$ from element $n$. Summing over $m$, using $dQ_n^{(m)} = -dQ_m^{(n)}$, immediately results in Eq. (B.41), the temperature increment equation. The purpose behind this simple GMM analysis is to provide a convenient starting point for adding new physics such as damage into the analysis.

## Acknowledgements

This work was funded by the DOE/DoD Joint Munitions Program (JMP), with Program Manager Eric Mas. Discussions with P. Rae, D. Dattelbaum, E. B. Orler, E. Brown, J. N. Johnson and F. Addessio are gratefully acknowledged.